\documentclass[journal, twoside]{IEEEtran}
\usepackage{amsmath,amsfonts}
\usepackage{algorithmic}
\usepackage{algorithm}
\usepackage{array}
\usepackage[caption=false,font=normalsize,labelfont=sf,textfont=sf]{subfig}
\usepackage{textcomp}
\usepackage{stfloats}
\usepackage{url}
\usepackage{verbatim}
\usepackage{graphicx}
\usepackage{cite}
\usepackage{booktabs,amsmath}
\usepackage{stfloats}

\usepackage{caption}
\usepackage{makecell}
\usepackage{multirow}
\usepackage{soul}
\usepackage{pgfgantt, tikz, rotating}
\usepackage[acronym,toc]{glossaries} 
\newacronym{IP}{IP}{Internet Protocol}
\newacronym{SP}{SP}{Security and Privacy}
\newacronym{IPSec}{IPSec}{Internet Protocol Security}
\newacronym{TCP}{TCP}{Transmission Control Protocol}
\newacronym{TCP/IP}{TCP/IP}{Transfer Control Protocol/Internet Protocol}
\newacronym{TLS}{TLS}{Transport Layer Security}
\newacronym{UDP}{UDP}{User Datagram Protocol}
\newacronym{CoAP}{COAP}{Constrained Application Protocol}
\newacronym{ESP}{ESP}{Encapsulation Security Payload}
\newacronym{AH}{AH}{Authentication Header}
\newacronym{SA}{SA}{Security Associations}
\newacronym{ISAKMP}{ISAKMP}{Internet Security Association and Key Management Protocol}
\newacronym{NAT}{NAT}{Network Address Translation}
\newacronym{SSL}{SSL}{Secure Sockets Layer}
\newacronym{HMAC}{HMAC}{Hash Message Authentication Code}
\newacronym{CA}{CA}{Certification Authority}
\newacronym{MAC}{MAC}{Media Access Control}
\newacronym{NAP}{NAP}{Network Attachment Point}
\newacronym{NFV}{NFV}{Network Functions Virtualization}
\newacronym{VNF}{VNF}{Virtualized Network Function}

\newacronym{SDN}{SDN}{Software Defined Networking}
\newacronym{ICN}{ICN}{Information-Centric Networking}
\newacronym{CDN}{CDN}{Content Delivery Networking}
\newacronym{DONA}{DONA}{Data-Oriented Networking Architecture}
\newacronym{PSIRP}{PSIRP}{Publish-Subscribe Internet Routing Paradigm}
\newacronym{PURSUIT}{PURSUIT}{Publish-Subscribe Internet Routing Paradigm}
\newacronym{PARC}{PARC}{Palo Alto Research Center}
\newacronym{NSF}{NSF}{National Science Foundation}
\newacronym{CCN}{CCN}{Content Centric Networking}
\newacronym{NDN}{NDN}{Named Data Networking}
\newacronym{IoT}{IoT}{Internet of Things}
\newacronym{DTN}{DTN}{Delay Tolerant Networking}
\newacronym{VANET}{VANET}{Vehicular Ad-hoc Networks}
\newacronym{FIA}{FIA}{Future Internet Architectures}

\newacronym{DoS}{DoS}{Denial of Service}
\newacronym{DDoS}{DDoS}{Distributed Denial of Service}
\newacronym{ACL}{ACL}{Access Control Lists}
\newacronym{PKI}{PKI}{Public Key Infrastructure}
\newacronym{NDO}{NDO}{Named Data Objects}
\newacronym{NRS}{NRS}{Name Resolution Service}
\newacronym{DNS}{DNS}{Domain Name Services}
\newacronym{URL}{URL}{Uniform Resource Locator}
\newacronym{HTTP}{HTTP}{Hyper Text Transfer Protocol}
\newacronym{HTTPS}{HTTPS}{Hyper Text Transfer Protocol over Secure Socket Layer}
\newacronym{PIT}{PIT}{Pending Interest Table}
\newacronym{CS}{CS}{Content Store}
\newacronym{FIB}{FIB}{Forwarding Information Base}

\newacronym{RF}{RF}{Rendezvous Function}
\newacronym{RP}{RP}{Rendezvous Point}
\newacronym{FF}{FF}{Forwarding Function}
\newacronym{FN}{FN}{Forwarding Node}
\newacronym{TF}{TF}{Topology Formation and Management Function}
\newacronym{TM}{TM}{Topology Management}
\newacronym{NSN}{NSN}{Name System Nodes}
\usepackage{enumitem}
\usepackage[]{todonotes}
\usepackage{cleveref}
\usepackage{balance}
\usepackage{threeparttable}
\usepackage{vcell}

\newcommand{\pie}[1]{%
\begin{tikzpicture}
 \draw (0,0) circle (1ex);\fill[rotate=90] (1ex,0) arc (0:#1:1ex) -- (0,0) -- cycle;
\end{tikzpicture}%
}
\usepackage{amssymb}
\usepackage{pifont}
\newcommand{\xmark}{\ding{55}}%

\newcommand{\ebtodo}[1]{{\todo[size=\tiny,color=pink]{\textbf{Bardhi:} #1}}}

\begin{document}

\title{Security and Privacy of IP-ICN Coexistence: \\ A Comprehensive Survey}

\author{Enkeleda~Bardhi\IEEEauthorrefmark{1},
        Mauro~Conti\IEEEauthorrefmark{2},
        Riccardo~Lazzeretti\IEEEauthorrefmark{1},
        Eleonora~Losiouk\IEEEauthorrefmark{2}
        
        \IEEEauthorrefmark{1}Department of Computer, Control and Management Engineering, Sapienza University of Rome\\
        \IEEEauthorrefmark{2}Department of Mathematics, University of Padua
        
        }

\markboth{Journal of Communication Surveys and Tutorials}{Bardhi \MakeLowercase{\textit{et al.}}: Security and Privacy of IP-ICN Coexistence: A Comprehensive Survey}


\maketitle
\begin{abstract}
%
Today's Internet is experiencing a massive number of users with a continuously increasing need for data, which is the leading cause of introduced limitations among security and privacy issues.
To overcome these limitations, a shift from host-centric to data-centric is proposed, and in this context, \gls{ICN} represents a promising solution.
Nevertheless, unsettling the current Internet's network layer -- i.e., \gls{IP} -- with \gls{ICN} is a challenging, expensive task since it requires worldwide coordination among Internet Service Providers (ISPs), backbone, and Autonomous Services (AS).
Therefore, researchers foresee that the replacement process of the current Internet will transition through the coexistence of IP and ICN.
In this perspective, novel architectures combine \gls{IP} and \gls{ICN} protocols.
However, only a few of the proposed architectures place the security-by-design feature.
Therefore, this article provides the first comprehensive \gls{SP} analysis of the state-of-the-art IP-ICN coexistence architectures by horizontally comparing the \gls{SP} features among three deployment approaches -- i.e., overlay, underlay, and hybrid -- and vertically comparing among the ten considered \gls{SP} features. 
Lastly, the article sheds light on the open issues and possible future directions for IP-ICN coexistence.
Our analysis shows that most architectures fail to provide several SP features, including data and traffic flow confidentiality, availability, and anonymity of communication.
Thus, this article shows the secure combination of current and future protocol stacks during the coexistence phase that the Internet will definitely walk across.

\end{abstract}

\begin{IEEEkeywords}
Security \& Privacy, Information-Centric Networking, Internet Protocol, IP-ICN coexistence
\end{IEEEkeywords}

\section{Introduction}
\label{ssec:intro}

The long journey of the Internet, designed to enable information sharing between a small group of researchers, started in the 1960s under the name ARPANET. Today’s Internet started officially in 1983 with the launch of \gls{TCP/IP} as a new communication protocol that allowed different networks to communicate.
In the last ten years, the Internet has been facing a massive change due to the increasing number of users, several devices used for various purposes, and the need for connectivity everywhere and anytime. According to Cisco~\cite{cisco}, 5.3 billion people worldwide used the Internet by 2021, representing 66\% of the world’s total population compared to 51\% up to 2018. 
From the same statistics, by 2023, each user will have 3.6 networked devices and connections, while up to 2018, each had 2.4 networked devices.
Considering this growing trend, the misalignment between the Internet’s initial and current usage model is becoming more prominent, highlighting several limitations. Such limitations include the availability of unique IP addresses, performance degradation, and \acrfull{SP} issues. 
To mitigate the former limitation, researchers proposed to switch from IPv4 to IPv6 protocol, going from 32 to 128 bits allocated for addressing purposes. Another mitigation was the introduction of \gls{NAT}~\cite{tsirtsis2000network} that maps different private addresses of devices located in a private network to a single public address through the presence of a firewall. 
Instead, the performance degradation of the current Internet is related to an ever-increasing number of users and devices used by each of them and their type of traffic. According to the Cisco Visual Networking Index~\cite{cisco1}, IP video traffic has grown three-fold from 2016 to 2021, reaching 227.6 Exabytes/month, while in 2016, it reached 70.3 Exabytes/month. Lastly, due to the lack of security by design, the Internet’s original design fails to provide some requirements---i.e., data confidentiality, integrity, and availability. The evolution of Internet Protocol (IP) to \gls{IPSec} or \gls{TLS} was introduced to handle the \gls{SP} issues found on the Internet over time. 
\par Considering these findings, both Academia and Industry agreed upon designing a new Internet architecture that shifts from an endpoint-based -- i.e., address hosts through their \gls{IP} addresses -- to a content-based communication---i.e., refer to contents by their names.
\acrfull{ICN} is the most research-targeted among the proposed paradigms, and it introduces the following benefits: (i) scalable and efficient data distribution by naming data instead of referring to its location~\cite{ambrosin2018security}; (ii) improved network load and communication latency due to the presence of routers caches~\cite{seetharam2017caching, ioannou2016survey}; (iii) security by design~\cite{tourani2017security, fu2018information, abdallah2015survey}; and (iv) enhanced support for mobility~\cite{fang2018survey, carofiglio2015scalable}.
Given all the benefits that \gls{ICN} brings, the transition towards it would be desirable. Nevertheless, the replacement of \gls{IP} with \gls{ICN} can take time due to \gls{IP} pervasiveness. Indeed, previous technology transitions include an intermediate phase during which the old and the new technology cohabit. In this context, the replacement of \gls{IP} will evolve through a transition phase during which both \gls{IP} and \gls{ICN} protocols will coexist. 
\gls{ICN} as a clean slate has been the focus of many research works, including several surveys. In~\cite{ioannou2016survey}, authors survey the caching feature of \gls{ICN}, focusing on the caching issues and policies.
Zhang et al.~\cite{zhang2013caching} presents a comprehensive survey on the techniques that reduce cache redundancy and improve cached content availability.
\gls{SP} properties of \gls{ICN} architectures have been surveyed in~\cite{tourani2017security}, where authors summarize security attacks and further elaborate on their impact on different \gls{ICN} features, including naming, caching, and routing. From an \gls{SP} point of view, Abdallah et al.~\cite{abdallah2015survey} provide a survey on attacks unique to \gls{ICN} architectures and a taxonomy of such attacks. In~\cite{xylomenos2013survey}, the authors study and survey \gls{ICN} architectures, identifying their core functionalities and weaknesses. Tortelli et al.~\cite{tortelli2016icn} survey the software tools used to experiment with \gls{ICN} architectures. 
%
%
%
\par Conversely from the clean-slate approach, the research on the coexistence of the future Internet paradigm -- i.e., ICN -- and the current one -- i.e., IP -- has been more limited~\cite{conti2020road,amadeo2016information,arshad2018recent,mars2019using}. 
Most of them address the IP-ICN coexistence from a performance point of view, deployment approaches and scenarios~\cite{conti2020road,rahman2018deployment}, and the combination of ICN with other technologies---e.g., \gls{SDN}, \gls{CDN}, \gls{NFV}, \gls{IoT}, 5G and \gls{VANET}.
However, none considers the security and privacy aspects of the IP-ICN coexistence. 
%
%
From the \gls{SP} perspective, the issues of ICN and IP protocol, taken singularly, have been widely studied, while such aspects in IP-ICN coexistence have not received attention.
Indeed, the deployment of \gls{ICN} in conjunction with the \gls{IP} differs from the clean-slate approach not only in terms of performance but also from the \gls{SP} perspective.
Here, unprotected communication between two heterogeneous coexistence network environments provides the potential for adversaries that cannot be detected by existing security mechanisms in standalone protocols~\cite{RamezanpourJJ23}.
The proposed security systems for the standalone protocols are designed to observe and respond to host- or content-based activities.
Therefore, the blended communication between such models without a security mechanism allows the adversary to exploit existing known attacks with less effort and higher intensity.
Nevertheless, the \gls{SP} study of such coexistence is challenging, mainly due to the lack of real-world deployment and open-source code of the proposed coexistence architectures. 
We here propose the first article that provides the \gls{SP} analysis of proposed IP-ICN coexistence architectures based on ten \gls{SP} features. 

\textbf{Contributions.} The main contributions of the paper are therefore listed as follows:
\begin{itemize}
    \item We reassess the traditional definition of the SP features by considering them in the new scenario of coexistence between IP and ICN protocols. 
    \item We provide the first complete and exhaustive \gls{SP} analysis of 20 architectures that address IP-ICN coexistence by considering the defined SP features.
    We group and then analyse the architectures according to their deployment approach---i.e., overlay, underlay, and hybrid. After that, we provide an in-group comparison of the analysed architectures while identifying the SP features that are not ensured. 
    %
    %
    \item We provide a discussion of the open issues in terms of not achieved SP features has been provided. Lastly, we accurately provide insights into the lessons learned and future directions for improving the \gls{IP}-\gls{ICN} coexistence in terms of security and privacy.
\end{itemize}
\textbf{Organization:} 
Section~\ref{sec:background} provides an overview of \gls{IP} and \gls{ICN} protocols, also focusing on their security models. Furthermore, we introduce different technologies -- i.e., \gls{SDN}, \gls{CDN}, \gls{NFV}, and \gls{DTN} -- exploited by the analysed coexistence architectures. 
Section~\ref{ssec:related-work} describes the state-of-the-art survey articles that address IP-ICN coexistence.
Section~\ref{sec:evaluation-parameters} introduces and motivates the features we select for the SP analysis.
Subsequently, we present the extensive SP analysis of 20 coexistence architectures according to their deployment approach--- i.e., overlay in~\Cref{sec:overlay}, underlay in~\Cref{sec:underlay} and hybrid in~\Cref{sec:hybrid}. We first describe each category, analyse ten SP features, and compare the architectures within a deployment approach according to the ensured SP features.
Section~\ref{sec:discussion} summarizes our findings while discussing open issues, lessons learned, and future directions. Finally, we conclude in Section~\ref{sec:conclusions}. 
%

\section{Overview of IP, ICN, and additional technologies}
\label{sec:background}

In this section, we first focus on \acrfull{IP} in~\Cref{ssec:ip} and \acrfull{ICN} in~\Cref{ssec:icn} protocols, as the pillars of the IP-ICN coexistence, by illustrating the protocol details together with its SP features. 
In addition, in~\Cref{ssec:additional-technologies}, we provide the same overview for the additional architectures -- i.e., \acrfull{SDN}, \acrfull{CDN}, \acrfull{NFV} and \acrfull{DTN} -- due to their high occurrence in IP-ICN coexistence architectures.
%
%
%
%
Table~\ref{tab:background} summarizes the general features, deployment type, and the key \gls{SP} features provided by design for each of the described protocols.
\begin{table*}[!ht]
    \caption{Overview of the protocols that are used to enable IP-ICN coexistence. The table summarizes the type of the current deployment of the described protocols -- i.e., deployment type --, their general characteristics regarding the communication model and other protocol-oriented attributes --i.e., general features --, and the security and privacy features guaranteed by design---i.e., SP features.}
    \centering
        \begin{tabular}{ c c  c c c c c}
            \toprule
            \multirow{2}{*}{\makecell{\textbf{Protocol}}} & \multirow{2}{*}{\textbf{Deployment Type}} & \multirow{2}{*}{\textbf{General Features}} & \multicolumn{4}{c}{\textbf{SP Features}} \\
            \cmidrule{4-7}
            & & & \makecell{Data Origin \\ Authentication} & \makecell{Data \\ Integrity} & \makecell{Data \\ Confidentiality} & \makecell{Consumer \\ Anonymity} \\ 
            \midrule
            \makecell{IP/IPSec} & \makecell{Clean slate} & \makecell{Host-based communication \\ IPv4 \& IPv6 \\ IPSec: AH and ESP} 
            & \multirow{5}{*}[-5em]{\parbox{2cm}{\centering Message Authentication Code (MAC) verification via shared secret keys for each packet}}
            & \multirow{5}{*}[-6em]{\parbox{2cm}{\centering AH for entire packet integrity, while ESP for packet headers integrity only}}
            & \multirow{5}{*}[-4em]{\parbox{2cm}{\centering In transport mode, ESP only encrypts the IP packet payload, while tunnel mode extends confidentiality to the entire encapsulated IP packet, thus payload and header}}
            & \multirow{5}{*}[-6em]{\xmark}\\
            \cmidrule{1-3}
            \makecell{SDN} & \makecell{Overlay in IP} & \makecell{Host-based communication \\ Centralized network\\ Control and data plane decoupling \\ Improved network management} & & & & \\
            \cmidrule{1-3}
            \makecell{CDN} & \makecell{Overlay in IP} & \makecell{Host-based communication \\ Efficient content distribution \\ Improved content availability} & & & &  \\
            \cmidrule{1-3}
            \makecell{NFV} & \makecell{Overlay in IP} & \makecell{Host-based communication \\ Hardware and software decoupling \\ Flexible and scalable networks} & & & &\\
            \cmidrule{1-3}
            \makecell{DTN} & \makecell{Overlay in IP} & \makecell{Host-based communication \\ Delay- tolerant applications} & & & &\\
            \midrule
            \makecell{ICN} & \makecell{Clean slate} & \makecell{Content-based communication \\ Content-based security \\ In-network caching \\ Name-based forwarding} 
            & \parbox{2cm}{\centering Data packets are signed by the producers, while interest packets are not signed by the consumers}
            & \parbox{2cm}{\centering Data packet integrity can be checked for integrity through signature verification, while for interest packets it is not applicable}
            & \parbox{2cm}{\centering \xmark}
            & \parbox{2cm}{\centering The consumer related information --e.g., IP address -- is not used given the data-centric nature of ICN} \\
            \bottomrule
        \end{tabular}
    \label{tab:background}
\end{table*}
%

%

\subsection{Internet Protocol (IP)}
\label{ssec:ip}
\acrfull{IP} is the core of today's Internet that over three decades ago, and it is also known as the TCP/IP architecture due to the presence of its two fundamental protocols---i.e., \gls{TCP} and \gls{IP}. 
Internet's ``hourglass" shape incurs due to the \gls{IP} in the network layer since the communication must navigate through IP regardless of the transport protocol, application, or network medium.
IP is the protocol in charge of forwarding IP datagrams during the host-based communication from the source counterpart toward the destination. For this purpose, each network entity is equipped with an identifier -- i.e., an IP address -- and an IP datagram containing both source and destination IP addresses in conjunction with the payload field that carries the actual data. The routers maintain the \gls{FIB} to define the next hop for each received packet and forward it toward its destination. 
While the Internet kept increasing, IP addressing space diminished, which led to the implementation of two new versions---i.e., IPv4 and IPv6. These two new versions created more addressing availability.  
\par \gls{IP} is not designed with the security-by-design, facing lots of \gls{SP} issues during its journey. The \gls{IPSec} suite~\cite{kent1998security, atkinson1995security} is considered the most advanced effort in standardizing IP security. \gls{IPSec} covers both versions of IP -- i.e., IPv4 and IPv6 -- and provides some basic security services, including IP datagrams' confidentiality, integrity, and origin authentication. To ensure the origin authentication and data integrity, \gls{IPSec} makes use of \gls{AH} protocol~\cite{ah}. Instead, to provide the full security triangle -- i.e., origin authentication, data integrity, and confidentiality---it uses the \gls{ESP} protocol~\cite{esp}. Furthermore, \gls{IPSec} uses the \gls{SA} to establish shared security attributes, including keys and algorithms. The \gls{AH} and \gls{ESP} protocols operate in transport mode or tunnel mode. In the former, the security mechanisms are applied to the upper layers' data -- i.e., packet payload -- and the IP header is left unprotected. Conversely, in the latter, the entire IP packet is protected by encapsulating it into a new datagram with a new outer IP header.

\subsection{Information-Centric Networking (ICN)}
\label{ssec:icn}

Information-Centric Networking (ICN) is a broad concept presented for the first time in 2001 as part of the TRIAD project~\cite{cheriton2000triad} to substitute the IP layer with a content layer. 
With such substitution, the purpose is to provide a better-performing infrastructure for content distribution and mobility.
Since its first proposal, different architectures adhering to the \gls{ICN} concept have been proposed. 
In 2010, \gls{NSF} presented a branch of the \gls{CCN} project called \gls{NDN}~\cite{ndn}. \gls{NDN} is considered the main project that is moving the future internet idea forward~\cite{icnrg}. 
In this article, we use \gls{ICN} and \gls{NDN} terms interchangeably.

\par \gls{NDN} relies on functionalities such as content naming, name-based routing, in-network caching, and content-based security~\cite{xylomenos2013survey, ahlgren2012survey}. 
The content in \gls{NDN} is recognized through a location-independent unique name that adheres to a hierarchical scheme. Such names are used during routing and forwarding. Therefore, \gls{ICN} gets rid of \gls{IP} addresses.
During routing and forwarding, the routers can store the data being forwarded in the cache and use it to satisfy future requests for the same content. 
Conversely,  \gls{NDN} tightly relate security concepts to the content itself and the use of trusted entities or cryptographic keys to achieving content-based security~\cite{smetters2009securing}. Here, content naming structure impacts content-based security~\cite{ghodsi2011naming}. By changing the security model to content-based, NDN ensures data origin authentication and integrity while keeping consumers' anonymity.
Indeed, human-readable names require a third party -- i.e., a trusted agent -- to verify that the delivered data corresponds to the requested content name. Instead, for flat content names, self-certification can be applied to them. Even in this case, a trusted third party must map the human-readable names to flat names. 
Additionally, in-network caches and the routing by name might raise the issues of \gls{SP}. Abdallah et al.~\cite{abdallah2015survey} classify the ICN vulnerabilities based on four possible targets: naming, caching, routing, and other miscellaneous attacks. Similarly, Tourani et al.~\cite{tourani2017security} make the same classification adding also \gls{DoS}~\cite{afanasyev2013interest, gasti2013and, compagno2013poseidon, dai2013mitigate, compagno2012ndn} as a severe attack related to availability, and application security~\cite{burke2013securing, burke2014secure, ambrosin2014covert}. In~\cite{tourani2017security}, the authors explore ICN's security, privacy, and access control concerns, covering user, cache, content privacy, traffic flow, confidentiality, communication anonymity, and access control. The authors in~\cite{ambrosin2018security} provide an analysis on two sides---i.e., network layer and resolution services. Instead, in~\cite{chaabane2013privacy}, the authors analyze CCN focusing on cache, content, name, and signature privacy. In~\cite{gasti2018content}, the authors classify the \gls{ICN} security issues based on two macro-categories: entity and router issues. Under these two categories, the authors elaborate on several issues raised in \gls{ICN}, including cache privacy, access control based on content, anonymous communication, and secure routing. Yu et al.~\cite{yu2018content} surveyed content protection challenges in \gls{ICN}, outlining security issues based on content name, availability, data integrity, and trust management.

\subsection{Additional Technologies}
\label{ssec:additional-technologies}
Here, we present an overview of the additional technologies that are used to enable IP-ICN coexistence. 
\subsubsection{Software Defined Networking (SDN)~\cite{SDNSurvey}} \label{sssec:sdn}
is an overlay technology implemented mainly in IP networks that separates the network activities into the control plane -- i.e., routing, name resolution -- and data plane---i.e., forwarding, storing, and caching. With such separation, \gls{SDN} introduces simpler network management and policy enforcement~\cite{kreutz2014software} than the traditional IP network architecture. 
The controller is a fundamental component in \gls{SDN} as it owns a broad view of the forwarding devices -- i.e., switches -- which are mainly configured via a programming interface. The most well-known is OpenFlow~\cite{mckeown2008openflow}. The controller maintains and manages the flow tables of the switches, which in turn can operate as switches, routers, firewalls, or \gls{NAT}. A switch can have one or more flow tables. Whenever the incoming traffic matches a rule installed by the controller, the switch performs specific actions---e.g., forwarding, dropping, and modifying.
%
\gls{SDN} networks follow the same security model as \gls{IP} networks. Besides, SDN introduces new assets and resources---i.e., flow tables on switches, the controller, the communication channel between the controller and switches, and the interface the controller uses to communicate with higher-level applications. 
Such resources might create new \gls{SP} challenges for the \gls{SDN} security model. For example, on the one hand, the controller's presence with a global view of the network offers some advantages in SDN, such as intrusion detection capabilities, malicious switch detection, or accountability~\cite{SDNSecurityProsCons}. 
On the other hand, its presence has been exploited by the attackers to launch \gls{DoS} and \gls{DDoS} attacks~\cite{yan2015software, yan2015distributed}. 
Similarly, many rules in the flow tables are placed on the switches that can be exploited for the same purpose. Despite the security issues and challenges that \gls{SDN} faces, several solutions have been proposed for both control and data planes~\cite{SDNSecurity, ahmad2015security}.
Several surveys address SP issues in SDN networks~\cite{ahmad2015security,alsmadi2015security,rahouti2022sdn,SDNSecurityProsCons,eliyan2021and}.

\subsubsection{Content Delivery Networking (CDN)~\cite{vakali2003content}}\label{ssec:cdn}
is deployed as an overlay on the Internet, aiming to overcome some of its limitations. 
\gls{CDN} improves availability, accessibility, and content distribution~\cite{pathan2007taxonomy, peng2004cdn} of IP networks. A CDN network consists of several caching nodes -- i.e., edge servers or surrogates -- that replicate content from the original servers and are strategically placed throughout the Internet. 
In particular, \gls{CDN} typically hosts static content, including images, video, media clips, advertisements, and other embedded objects for dynamic web content. 
Furthermore, \gls{CDN} can be a centralized, hierarchical infrastructure under specific administrative control or a completely decentralized system.
It can be built through one of the following approaches: (i) overlay model or (ii) network model. Several servers and caches at different network locations handle the distribution of specific content types in the former. The network components can recognize the application types and apply different forwarding policies in the latter.
%
CDN inherits the IP security model. New networking components such as edge servers, caches, and routing mechanisms also introduce new \gls{SP} issues~\cite{ghaznavi2021content}.
Similarly to \gls{IP} and \gls{ICN} networks, \gls{CDN} can be the target of well-known attacks such as \gls{DoS}, cache pollution~\cite{mubarok2014lightweight},  cache poisoning~\cite{nguyen2019your}.

\subsubsection{Network Functions Virtualization (NFV)~\cite{han2015network}}\label{ssec:nfv}
is a technology that enables the virtualisation of network functions~\cite{li2015software}. 
Indeed, \gls{NFV} enables different network functions in various locations---e.g., network edge, network nodes, or data centers.
It is usually used in conjunction with \acrshort{SDN}, but unlike the latter, \acrshort{NFV} copes only with network functions virtualization.
This virtualization process allows the export of network functions from the underlying hardware infrastructure to general software that runs on dedicated devices.
From the security point of view, the NFV Infrastructure (NFVI) should adopt standard security mechanisms for authentication, authorization, encryption, and validation~\cite{yang2016survey}. Furthermore, NFV technology faces management and orchestration security challenges ~\cite{keeney2014towards}.
In~\cite{alwakeel2018survey}, the authors categorise the NFV security issues into \gls{DoS}, infrastructure integrity threats, misuse of resources, malicious insiders, privileges modification, and confidentiality attacks based on shared resources.

\subsubsection{Delay Tolerant Networking (DTN)~\cite{burleigh2003delay}}\label{ssec:dtn} is an evolution of Mobile Ad-hoc Networks (MANET) and it represents a sparse and intermittently connected mobile ad-hoc network. 
In DTN, reliable communication and end-to-end connectivity are unavailable for message transmission. Therefore, this form of networking is suitable only for high-latency applications, where the latency may be in hours or days. The ``store and forward" approach adopted by DTN helps to increase message delivery probability irrespective of the time taken to deliver the message over the MANET. 
Menesidou et al.~\cite{menesidou2017cryptographic} discuss trust and cryptographic key management for DTN. 

\section{Related Work}
\label{ssec:related-work}
The research community lacks a contribution that targets the \gls{SP} analysis of IP-ICN coexistence. Thus, in this article, we aim to fill such a gap.
However, different surveys regarding IP-ICN coexistence have been proposed, as shown in \Cref{tab:related-work-surveys}.
%
Most proposed surveys target integrating ICN concepts in the \gls{IoT} networks. Indeed, due to their data-centric nature, IoT networks would benefit from several ICN features. 
In~\Cref{tab:related-work-surveys}, we compare our article with the other state-of-the-art surveys. For such comparison, we consider the protocols involved in the coexistence, the number of analysed architectures, if any, the aspects of coexistence in the survey, and the SP features considered.
\begin{table*}[!ht]
    \caption{State-of-the-art IP-ICN coexistence surveys and their characteristics.}
    \centering
    \resizebox{0.85\textwidth}{!}{
        \begin{tabular}{c c c c c}
            \toprule
            \multirow{2}{*}{\makecell{\textbf{Reference}}} & \multicolumn{4}{c}{\textbf{Features}} \\
            \cline{2-5} \\
            & \makecell{\textbf{Coexistence} \\ \textbf{Protocols}} & \makecell{\textbf{Number of Analysed} \\ \textbf{Architectures}} & \makecell{\textbf{Main Survey}\\ \textbf{Aspects}} & \makecell{\textbf{Security and Privacy}\\ \textbf{Features}} \\
            \midrule
            \cite{conti2020road} & ICN-IP & 14 & \makecell{Deployment approaches \\ Deployment scenario \\ Forwarding \\ Storage \\ Management}  & None \\
            \midrule
            \cite{amadeo2016information} & ICN-IoT & 1 & \makecell{ICN features in IoT} & None \\
            \midrule
            \cite{arshad2018recent} & ICN-IoT & 9 & \makecell{ICN features in IoT \\ Caching, Naming, Mobility \\ \& Security schemes \\ OS and Simulators} & \makecell{Device Security \\ Content Security \\ Device and content security} \\
            \midrule
            \cite{mars2019using} & ICN-IoT & 1 & \makecell{ICN features in IoT \\ ICN-IoT Applications} & None \\
            \midrule
            \cite{nour2019survey} & ICN-IoT &  5 & \makecell{ICN features in IoT \\ ICN-IoT Applications \\ Mobility \\ Quality of Service}  & \makecell{Authorization \& access control \\ Privacy}\\
            \midrule
            \cite{din2019review} & ICN-IoT & 1 & \makecell{ICN features in IoT \\ Integration of ICN-IoT} &  None\\
            \midrule
            \cite{aboodi2019survey} & ICN-IoT & 1 & \makecell{ICN features in IoT \\ ICN-based IoT applications \\ Access Control Mechanisms} & None \\
            \midrule
            \cite{djama2020information} & ICN-IoT & 15 & \makecell{Deployment approach \\ Service models \\ Infrastructure modes \\ Guaranteed QoS \\  OS and platforms} & None \\
            \midrule
            \cite{RahmanHKIDBK23} & ICN-IoT & 8 & \makecell{ICN-IoT-FL integration \\ Integration benefits \\ Security and Privacy \\ Challenges} & \makecell{Denial of Service \\ Access Control \\ Naming privacy}\\
            \midrule
            \cite{gur2020convergence} & ICN-5G & 1 & \makecell{ICN-MEC benefits, \\ challenges \& applications \\ Standardisation}  & None \\
            \midrule
            \cite{serhane2020survey} & ICN-5G & 1 & \makecell{ICN-5G integration \\ In-network caching schemes \\ Naming schemes}  & None \\
            \midrule
            \cite{khelifi2018security} & ICN-VANET & 1 &  \makecell{NDN integration in VANET \\ VANET attacks \\ NDN-VANET attacks}  & \makecell{Denial of Service \\ Black/Gray/Worm-hole attacks \\ Man-in-the-middle} \\
            \midrule
            \cite{amadeo2016nov} & ICN-VANET & 1 & \makecell{ICN-VANET integration \\ Application \\ Mobility}  & None \\
            \midrule
            \cite{wang2023towards} & ICN-Vehicular Cloud & 3 & \makecell{ICN-VC Features \\ ICN-VC challenges} & None \\
            \midrule
            \cite{fayyaz2023information} & \makecell{ICN-MANET \\ ICN-VANET} & 7 & \makecell{Producer Mobility \\ Consumer Mobility \\ Mobility Challenges} & None \\
            \midrule
            \cite{MusaZLP22} & ICN-IoV & 1 & \makecell{ICN-IoV integration \\ ICN-IoV Challenges} & None \\
            \midrule
            Our Survey & ICN-IP & 20 &  \makecell{Security and privacy analysis \\ Deployment Approaches \\ Deployment scenarios \\ Network layer features \\ Additional technologies} &  \makecell{Trust \\ Data origin authentication \\ Data integrity \\ Data confidentiality \\ Peer entity authentication \\ Accountability \\ Authorization \& access control \\ Availability \\ Traffic flow confidentiality \\ Anonymous communication} \\
            \bottomrule
        \end{tabular}}
    \label{tab:related-work-surveys}
\end{table*}
\par Amadeo et al.~\cite{amadeo2016information} surveyed the opportunities and challenges for integrating ICN in IoT networks. Here, the authors analyse the various benefits of ICN in IoT networks. Furthermore, the survey provides insights into ICN security in IoT without analysing the security and privacy model that must be considered in such an environment.
Arshad et al.~\cite{arshad2018recent} also targeted ICN-IoT coexistence. Here, this survey reviews ICN for IoT, including ICN models and their feasibility for IoT. Additionally, this article considers the caching techniques, naming schemes, mobility handling mechanisms, and operating systems and simulators. Here, it discussed content and device security schemes. However, they mainly focused on authorization, access control, and data integrity without considering other security features.
Mars et al.~\cite{mars2019using} surveyed the application of ICN in IoT networks. Here, they described various ICN-IoT applications and provided a comparison between them. However, the survey does not provide any security and privacy insights.
Nour et al.~\cite{nour2019survey} provided a comprehensive ICN-IoT survey. This article analyses the benefits of ICN in IoT networks while describing the ICN solutions for IoT according to the domain application. Moreover, the article discusses the mobility schemes of wireless IoT networks while the security considerations are only focused on existing works related to authorization, access control, and privacy.   
Din et al.~\cite{din2019review} surveyed the integration of ICN in IoT applications, going through all the design issues---e.g., naming, caching, scalability. The article additionally discusses the communication standards of IoT and the integration of ICN-based IoT with the existing architectures. However, the article completely lacks security and privacy considerations.
Another article that surveys the integration of ICN in IoT applications is proposed by Aboodi et al.~\cite{aboodi2019survey}. After providing a detailed analysis of ICN features, the article describes ICN-IoT-related problems---i.e., content and device naming, caching management, forwarding, and routing. The article additionally surveys access control mechanisms in ICN-IoT scenarios. Besides access control considerations, the paper lacks security and privacy analysis.
Djama et al.~\cite{djama2020information} survey ICN-IoT deployment approaches, service, and infrastructure models. Furthermore, the article also surveys the supported features of ICN-IoT solutions and the main ICN-IoT operating systems and platforms. Lastly, it also analyses service quality while it lacks security and privacy analysis. 
Rahman et al.~\cite{RahmanHKIDBK23} surveyed the integration of various technologies -- i.e., IoT, Federated Learning and \gls{ICN} -- and listed the benefits from each integration. Additionally, the article discusses the security and privacy issues of such integration. However, the provided discussion focuses more on the known security issues of ICN and their influence on integration.
\par In their work, Conti et al.~\cite{conti2020road} provide an overview of state-of-the-art solutions that target IP-ICN coexistence. They classify the coexistence architectures based on their deployment approach and scenario. They also discuss the addressed coexistence requirements and possible additional technologies to facilitate such coexistence. Here, the authors also make preliminary considerations on security for some of the surveyed architectures. Nevertheless, proper security and privacy analysis are not provided.
Gur et al.~\cite{gur2020convergence} provide the challenges and opportunities in deploying ICN in the 5G networks. In particular, the article targets Multi-access Edge Computing (MEC), a crucial technology that enables 5G requirements. Here, the paper investigates the integration of ICN and MEC, focusing on the mutual benefits of such integration, ICN-MEC applications, and standardisation issues. Lastly, the article slightly mentions the security and privacy issues emerging from such integration without providing an analysis.
Serhane et al.~\cite{serhane2020survey} survey the in-network caching and content naming in 5G-enabled ICN networks. The article analyses the ICN paradigm's applicability and feasibility in the next-generation 5G networks. Moreover, the paper surveys such networks' content naming and in-network caching while highlighting such integration's challenges and future directions.
Khelifi et al.~\cite{khelifi2018security} provides an overview of security and privacy issues in Vehicular Named Data Networks. This article surveys the existing  Vehicular Ad-hoc Networks (VANET) attacks and how NDN can deal with such attacks. Although a valid tentative to summarize the known attacks in such an environment, the proposed article needs more detailed security and privacy analysis according to the integration of NDN and VANET.
Amadeo et al.~\cite{amadeo2016nov} surveyed the information-centric vehicular networks, focusing on the state-of-the-art proposals that extend the ICN paradigm to accommodate VANET peculiarities. The article argues the advantages of integration of ICN in current vehicular networks and points to future research possibilities in this direction. However, the article lacks security and privacy analysis. 
Wang et al.~\cite{wang2023towards} provided an overview of the Vehicular Cloud (VC) by studying state-of-the-art VC architectures and especially, the \gls{ICN}-based VC. In particular, the article presents the differences between a vanilla vehicular network and a vehicular cloud while also elaborating on the open challenges of VC. Nevertheless, the article lacks security and privacy analysis.
Fayyaz et al.~\cite{fayyaz2023information} surveyed the mobility issues faced by \gls{ICN} and the provided efforts to overcome these issues. The article also discusses the challenges of integrating ICN-based mobility in environments such as Mobile Ad-Hoc Networks (MANET) and VANET. The article does not analyse security and privacy issues regarding ICN integration in such environments.
Lastly, Musa et al. \cite{MusaZLP22} reviewed the integration of ICN for the Internet of Vehicles (IoV) environment. Additionally, the article analyses the role of other enables -- i.e., Edge Computing, Artificial Intelligence, and Machine Learning -- in solving known IoV issues. Here, the article proposes an ICN-based IoV architecture while also discussing the shortcomings of such integration. Lastly, the authors discuss the security and privacy solutions for the known ICN issues. However, the article lacks an analysis of the issues faced in the ICN-IoV context.

\emph{Conversely from the state-of-the-art surveys, in this article, we aim to present the first comprehensive analysis of coexistence between ICN and IP protocols from the SP point of view. We study 20 state-of-the-art coexistence architectures considering ten SP features to fulfill this aim.}

\section{Coexistence Deployment and SP Features}
\label{sec:evaluation-parameters}

For evaluating the \gls{SP} properties of the coexistence architectures, we consider all the possible deployment approaches and scenarios of the ICN and IP protocol. Their description is provided in Section~\ref{ssec:deployment}. 
After that, in Section~\ref{ssec:features}, we present the set of ten \gls{SP} features mainly used to evaluate the security and privacy aspects of IP and ICN protocols.
Then, in Section~\ref{ssec:criteria}, we argue a new definition for each SP feature for the IP-ICN coexistence.
Lastly, in~\Cref{ssec:analyzed-architectures}, we briefly present the coexistence architectures we have analysed.

\subsection{Deployment approaches and scenarios}
\label{ssec:deployment}

In the IP-ICN coexistence scenario, one of the most critical issues concerns the deployment type of the \gls{ICN} protocol into the existing \gls{IP} infrastructure. 
Conti et al.~\cite{conti2020road} define three central deployment approaches in combining these two protocols---i.e., overlay, underlay, and hybrid. 
In the overlay approach, \gls{ICN} protocol is accommodated on top of \gls{IP}. Conversely, in the underlay approach, \gls{ICN} runs under \gls{IP} protocol, and in the hybrid approach, they both cohabit. 
Independently from the deployment approach, the main focus is connecting \gls{ICN} and \gls{IP} ``islands" through \gls{ICN} and \gls{IP} ``ocean ". Here, an ``island" can be a single or a group of devices, applications, or servers running either \gls{ICN} and \gls{IP} protocol. In contrast, an ``ocean" is a network containing components that run different architectures. 
Envisioning the presence of these ``islands" and ``oceans", the deployment scenarios for coexistence architectures are:
\begin{itemize}
    \item ICN ``islands" communicating through an IP ``ocean"
    \item IP ``islands" communicating through an ICN ``ocean"
    \item ICN and IP ``islands" communicating through an IP ``ocean"
    \item ICN and IP ``islands" communicating through an ICN ``ocean"
    \item ICN and IP ``islands" located in separate ``oceans" (i.e., border ``island")
\end{itemize}
Throughout the article, we consider the deployment approaches and scenarios while analysing each coexistence architecture. Indeed, the presence of heterogeneous ``islands" in either ICN or IP ``ocean" also impacts the security model of the considered architecture.

\subsection{Considered SP Features}
\label{ssec:features}

To evaluate how the proposed IP-ICN coexistence architectures cope with \gls{SP} challenges, we select ten representative features used in previous research works that target both the security and privacy of a specific protocol--- whether IP or ICN. These features are described in the Internet security glossary~\cite{shirey2007internet}, and we shortly describe them as follows:  

\begin{description}

 \item[Trust:] assurance -- sometimes based on inconclusive evidence -- that a) the system behaves as expected and according to the specifications and b) a trusted source provides the content.
    
    \item[Data integrity:] certainty that the data has not been modified or destroyed unauthorizedly.
    
    \item[Data origin authentication:] endorsement that the source of the received data is as claimed.
    
    \item[Data confidentiality:] assurance that data is not rendered available to unauthorized entities.
    
    \item[Peer entity authentication:] endorsement that a peer entity in an association is the one claimed.
    
    \item[Accountability:] certainty that the actions of an entity can be traced to make it responsible for its actions. 
    
    \item[Authorization \& access control:] assurance that a) a specific entity securely accesses -- e.g., reads, writes, deletes -- some resources and b) a specific resource is protected against unauthorized access. 
    
    \item[Availability:] certainty that a resource is usable, accessible, or operational upon demand by an authorized entity.
    
    \item[Traffic flow confidentiality:] a set of countermeasures to traffic analysis.
    
    \item[Anonymous communication:] assurance that communicating entities' identities can not be determined.
    
\end{description}

\subsection{SP Features in IP-ICN Coexistence}
\label{ssec:criteria}

The SP features described in~\Cref{ssec:features} up to date have been used to analyse single protocols, but for the context of IP-ICN coexistence, these features need to be revisited.
In such coexistence, the SP issues are mainly associated with the deployment scenario---i.e., the combination of islands and oceans.
Therefore, in this section, we describe the selected SP features considering the IP-ICN coexistence based on the previous deployment scenarios in~\Cref{ssec:deployment}. 
\emph{We emphasise that this revisiting of the SP features is preliminary and provides an understanding of the criteria used to analyse the coexistence architectures.} Table~\ref{tab:feature-coexistence} summarizes our considerations. Here, we categorize if a specific deployment scenario (i) is compliant with the current advancements of the IP and ICN protocols, taken singularly, or (ii) it is compliant only under modifications that need to be addressed during the design phase or (iii) it is too complex and needs a rethinking.
%
%
%
%

%
%
%
%
%
%
\begin{table*}[t]
    \caption{Revisited SP feature for the IP-ICN coexistence.}
    \centering
    \resizebox{0.87\textwidth}{!}{
        \begin{tabular}{ c  c  c  c c c}
            \toprule
            \multirow{2}{*}{\makecell{\textbf{Feature}}} & \multicolumn{5}{c}{\textbf{Deployment Scenarios for Coexistence}}\\
            \cline{2-6} \\
            & \makecell{\textbf{ICN islands in} \\ \textbf{IP ocean}} & \makecell{\textbf{IP islands in} \\ \textbf{ICN ocean}} & \makecell{\textbf{ICN and IP islands} \\ \textbf{in IP ocean}} & \makecell{\textbf{ICN and IP islands} \\ \textbf{in ICN ocean}} & \makecell{\textbf{Border island}} \\
            \midrule
            Trust & $\ast$ & $\ast$ & $\ast$ & $\ast$ & $\ast$ \\ 
            \midrule
            Data origin authentication & \checkmark & \checkmark & $\ast$ & $\ast$ & \xmark \\
            \midrule
            Peer entity authentication & $\ast$ & $\ast$ & $\ast$ & $\ast$ & $\ast$  \\
            \midrule
            Data integrity & \checkmark & \checkmark & $\ast$ & $\ast$ & \xmark \\
            \midrule
            Authorization \& access control & $\ast$ & $\ast$ & $\ast$ & $\ast$ & $\ast$ \\
            \midrule
            Accountability & $\ast$ & $\ast$ & $\ast$ & $\ast$ & $\ast$ \\
            \midrule
            Data confidentiality & \checkmark & \checkmark & \xmark & \xmark & \xmark \\
            \midrule
            Availability & \xmark & \xmark & \xmark & \xmark & \xmark \\
            \midrule
            Anonymous communication & $\ast$ & $\ast$ & \xmark & \xmark & \xmark \\
            \midrule
            Traffic flow confidentiality & \xmark & \xmark & \xmark & \xmark & \xmark \\
            \bottomrule
        \end{tabular}}
        \begin{tablenotes}
          \item[a] Legend: \checkmark - compliant to the current definition and mechanisms for IP and ICN protocols; $\ast$ - compliant to the current definitions and mechanisms for IP and ICN protocols only under certain criteria to be met in the design phase; \xmark - too complex to meet the current definitions and mechanisms for IP and ICN protocols.
        \end{tablenotes}
    \label{tab:feature-coexistence}
\end{table*}
\par Generally, we consider \textbf{trust} in \gls{IP} islands based on securing communication endpoints---e.g., hosts or networks. This trust model was designed to patch security issues in the current IP architecture. 
Conversely, for \gls{ICN} islands, trust is based on securing the content itself. Trust in content can be expressed at different levels of granularity---i.e., from securing the single content object to the entire namespace. 
However, in the native design of \gls{ICN}, trust management is unspecified and left to the application itself. 
%
Given the above considerations, trust is a design-dependent SP feature for all deployment scenarios, and as such, it is usually considered a partially fulfilled feature for coexistence architectures. %
\par \textbf{Data origin authentication} in \gls{IP} networks is ensured by \gls{IPSec} and in particular by \gls{AH} and \gls{ESP} protocols. In tunnel mode, the communication parties securely negotiate a shared secret key to generate a \gls{HMAC} for each packet~\cite{krawczyk1997hmac}. Instead, in \gls{ICN}, each content producer signs the data packets to provide data origin authentication. Therefore, consumers can validate the signature using the producer's public key before consuming the content. However, in this design, routers are not required to perform such validation since it is reputed as an expensive task. Furthermore, the consumers are not required to sign the interest packets even if such packets contain a small payload.
%
%
In the IP-ICN coexistence perspective, we argue that the above-described mechanisms can be applied in both oceans and islands for the first two deployment scenarios--i.e., ICN island in the IP ocean and IP islands in the ICN ocean. Instead, for the heterogeneous islands deployment scenario -- i.e., ICN and IP islands in IP or ICN ocean -- data origin authentication must also be guaranteed for the packets flowing from or to different islands. Therefore, since this challenge in these scenarios must be carefully addressed during the coexistence design phase, we mainly consider data origin authentication as partially fulfilled. Similarly, the border island -- i.e., heterogeneous islands and oceans -- represents a complex scenario that requires additional effort during the design to ensure signature verification procedures between islands and oceans. Thus, for this scenario, data origin authentication is mainly considered to need to be fulfilled during the architecture analysis.
A similar rationale is followed for the \textbf{data integrity} definition in the IP-ICN coexistence.
\par IPSec ensures \textbf{peer entity authentication} through \gls{ISAKMP} which is used for \gls{SA} establishment. Among several functionalities, \gls{ISAKMP} defines procedures for authenticating communicating peers. Here, certificates can bind the entity identity to the public key. However, using certificates requires a third party or \gls{CA} to manage the certificate creation, signing, and distribution. In ICN, peer entity authentication is not provided for consumers or producers and is left to the application. 
%
%
Therefore, peer entity authentication in the IP-ICN coexistence is feasible only if a proper entity authentication for ICN islands is adequately designed. To this extent, we fill peer entity authentication as design dependent for all the deployment scenarios, and we mainly consider it as partially fulfilled.
\par For the \textbf{authorization \& access control} we consider the \gls{ICN} islands unable to provide such feature. However, some solutions~\cite{fotiou2012access, singh2012trust} propose using additional authorities or secure communication channels by authenticating each content consumer. Another work~\cite{li2016attribute} proposes an Attribute-Based Encryption (ABE) based scheme to provide a security control mechanism in \gls{ICN}.
Instead, for the \gls{IP} islands, it is achieved through \gls{ACL}~\cite{qian2001acla}. However, these solutions also carry known vulnerabilities---i.e., address spoofing. 
Given these considerations, in the IP-ICN coexistence, it is expected that authorization and access control are design dependent for all the deployment scenarios. In this context, the coexistence architectures must properly shape the authorization and access control mechanism for ICN islands, especially for the border island scenario. Therefore, during the analysis, we mainly consider this feature as partially fulfilled.
\par The \textbf{accountability} feature depends on the presence of the parties' identities in the communication. \gls{IP} networks are the target of IP address spoofing attacks, and such vulnerability has been patched by applying egress filtering on routers or directly securing the communication channel---i.e., \gls{IPSec}. In \gls{ICN}, producers can guarantee accountability only by assuming that every content served by producers is signed. On the other hand, ICN can achieve consumer accountability if all the interest requests are signed, or the consumer's identity is included in the issued interest request, which is not part of the ICN's original design.
%
Considering the IP-ICN coexistence, accountability for ICN islands and oceans is design-dependent. As such, we expect considerations of adequate mechanisms to ensure this feature in the design phase, and we mainly consider this feature as partially achieved for the architecture analysis.
\par Furthermore, while analyzing \textbf{data confidentiality}, we observed that most architectures do not elaborate on the mechanisms used to achieve this feature. Indeed, assuring the confidentiality of packets flowing from different islands and oceans is not trivial.
Notwithstanding, the existing encryption schemes for IP and ICN networks can also be applicable in the IP and ICN islands. Therefore, ICN islands in the IP ocean and IP islands in the ICN ocean comply with the current IP and ICN protocol definitions. Instead, the other deployment scenarios urge a redefinition of the encryption and decryption mechanisms for packets flowing from different islands in heterogeneous oceans. For these considerations, we consider data confidentiality as needing to be fulfilled for most of the coexistence architectures.
\par In \gls{ICN} islands \textbf{anonymous communication} can be achieved due to the privacy-by-design nature of ICN protocol for the consumers. On the other hand, additional factors -- i.e., caches and routing tables -- might risk communication anonymity. Instead, in IP islands, communication anonymity can not be guaranteed. For this purpose, anonymity mechanisms -- e.g., TOR~\cite{tor} -- can be used.
In the IP-ICN coexistence context, in the heterogeneous islands and oceans scenarios -- i.e., ICN and IP islands in IP or ICN island and border island -- both the definition and the mechanisms for communication anonymity must be revisited, especially for the communication among different islands. Therefore, we consider this feature not fulfilled for the coexistence architectures adhering to these scenarios.
\par \textbf{Availability} and \textbf{traffic flow confidentiality} are among the massively not fulfilled features due to well-known issues for each of them in \gls{IP} and \gls{ICN} networks. Such issues are also transferred to the IP-ICN coexistence. Thus, we mainly consider this feature as not fulfilled by the coexistence architectures.

\subsection{Analyzed Architectures}
\label{ssec:analyzed-architectures}

\begin{figure*}[!ht]
\centering
\includegraphics[width=0.9\linewidth]{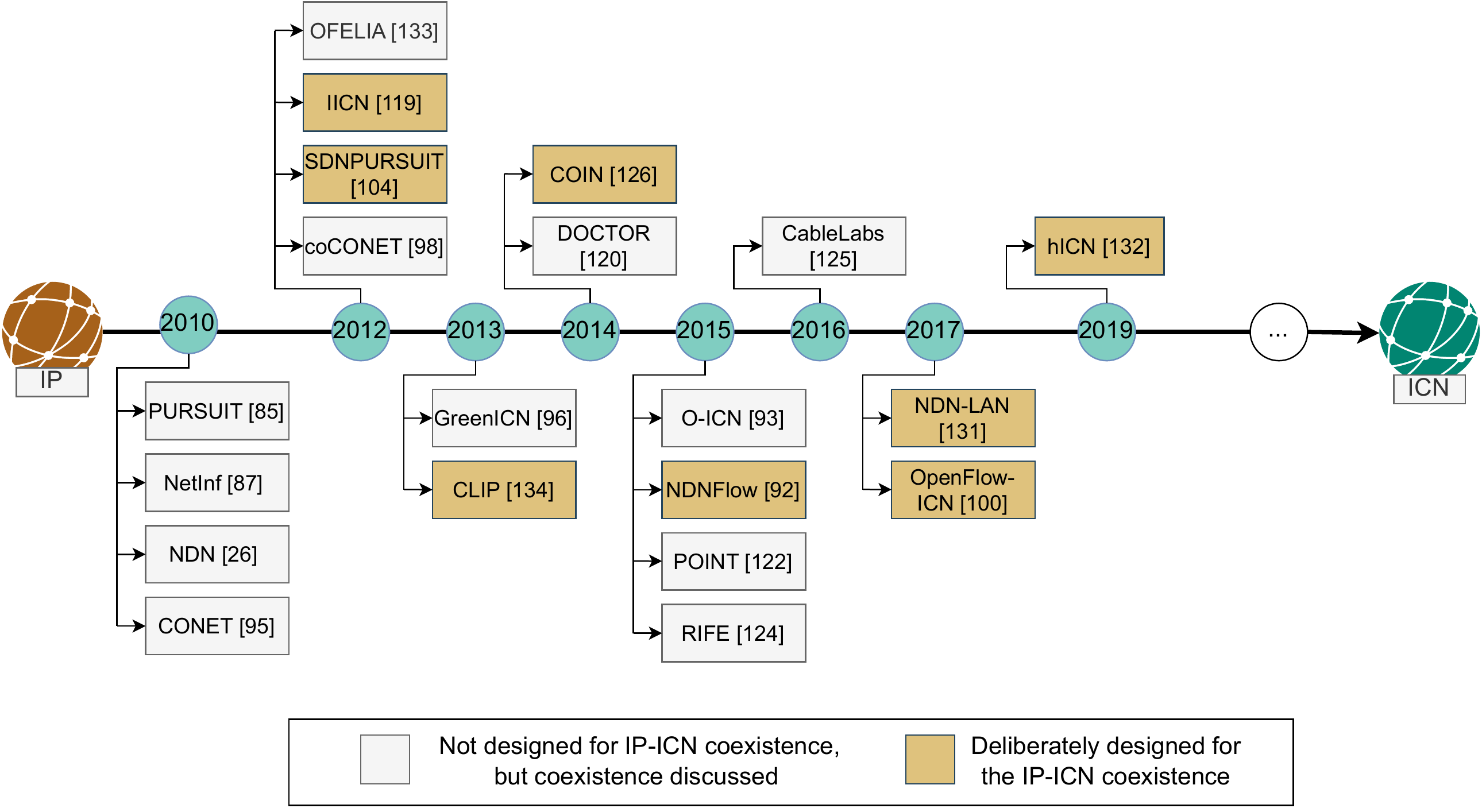}
\caption{Timeline of the analysed architectures starting from 2010. The analysis encompasses the architectures proposed as new ICN architectures -- i.e., gray boxes -- that further discuss their possible coexistence with the IP protocol. Furthermore, we analyse the architectures that deliberately design the IP-ICN coexistence---i.e., brown boxes.}
\label{fig:timeline}
\end{figure*}

\begin{figure*}[!ht]
\centering
\includegraphics[width=1\linewidth]{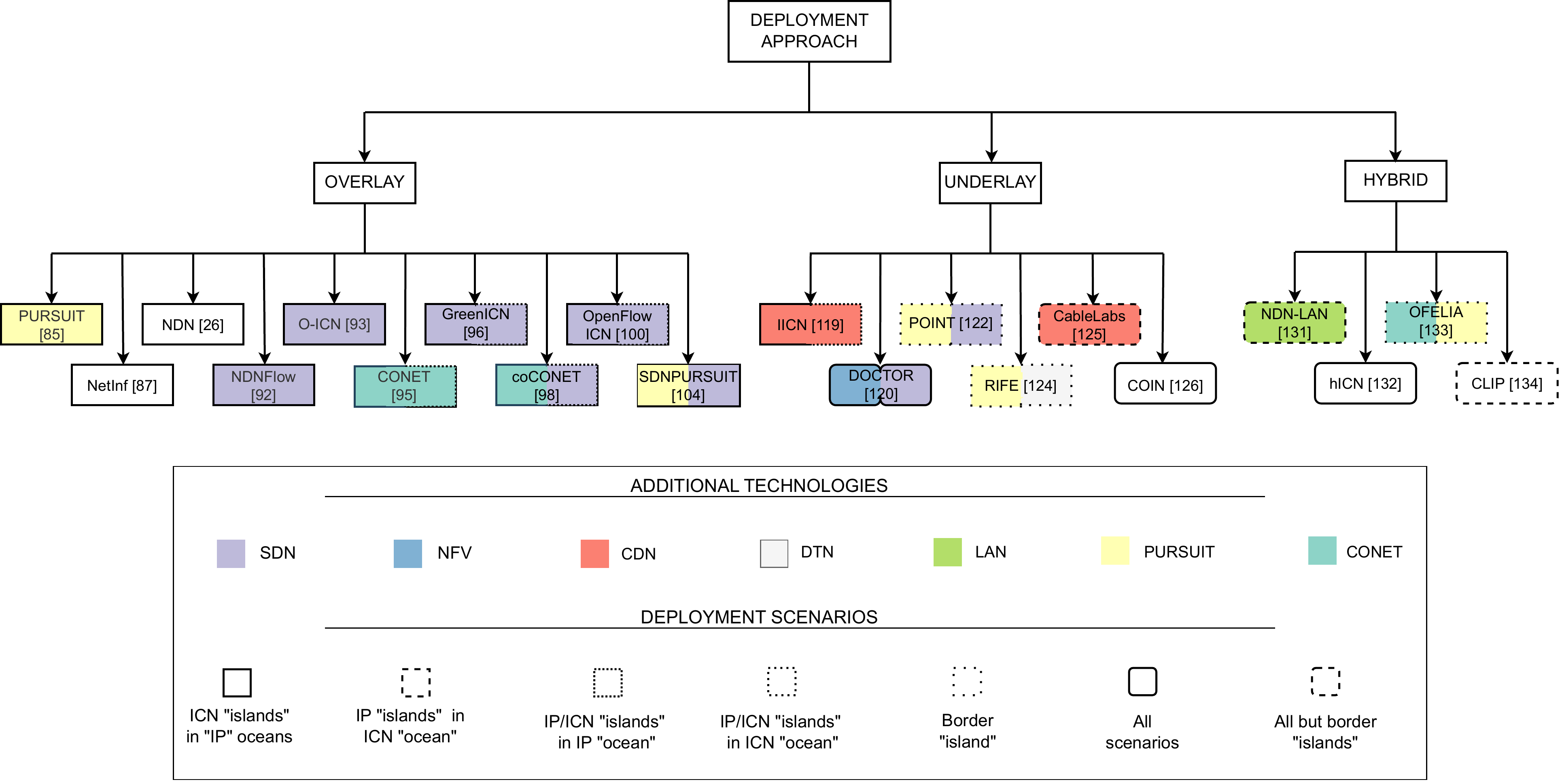}
\caption{The IP-ICN coexistence deployment approaches and the architectures adhering to each group. Furthermore, the chart highlights the deployment scenarios each architecture considers encoded in different lines. Lastly, the chart shows the inheritances from additional technologies used during the design of each architecture encoded in colours.}
\label{fig:chart}
\end{figure*}

This section briefly introduces the analysed architectures that address the IP-ICN coexistence.  
Overall, we have considered the architectures that deliberately address the coexistence and the clean-slate ICN architectures that discuss their possible coexistence with the IP protocol.
The analysed coexistence architectures are partitioned into three categories according to their deployment approach---i.e., overlay, underlay, and hybrid. 
Among 20 analysed architectures, ten adhere to the overlay deployment approach, while six to the underlay and four to the hybrid.
Fig.~\ref{fig:timeline} depicts the timeline of the analysed coexistence architectures.
As reflected in such a figure, Academia, and Industry mainly focused on the design of new \gls{ICN} architectures, notably after the launch of the ICN project in 2010. 
On the other hand, the architectures designed deliberately for IP-ICN coexistence are becoming numerous starting from 2017. 
Indeed, the current maturity of ICN and the known issues of IP are pushing the focus toward their coexistence.
Instead, Fig.~\ref{fig:chart} depicts the analysed coexistence architectures categorised based on the deployment approach. 
The color encoding represents the additional technologies used to enable the IP-ICN coexistence, while the line encoding shows the deployment scenarios. 
SDN is among the key enablers of such coexistence in terms of additional technologies, especially for the overlay deployment approach.
Indeed, SDN unlocks centralized control over the network that becomes fundamental when accommodating new -- i.e., ICN -- protocol semantics into existing -- i.e., IP -- protocol.
Lastly, regarding deployment scenarios, the overlay architectures mainly cover the ICN islands in the IP ocean scenario, given the placement of ICN on top of the IP protocol. 
Conversely, the underlay and hybrid architectures involve more scenarios, given the former approach's border gateways and the latter's dual switches. 
%
%
%
%

\section{Overlay Architectures}
\label{sec:overlay}
%
In this section, we analyse the overlay architectures, where the \gls{ICN} protocol runs on top of \gls{IP} to enable the communication between different \gls{ICN} islands through the existing IP infrastructure. 
We first shortly describe the architectures adhering to this category (\Cref{ssec:overlay-description}) focusing mostly on the communication paradigm, the node model and implementation particularities.
Then we provide a detailed analysis of the \gls{SP} features for each architecture (\Cref{ssec:overlay-sp-analysis}). 
Lastly, we provide a comparison between the analysed architectures (\Cref{ssec:overlay-comparison}).
%

\subsection{Description}
\label{ssec:overlay-description}
%
\subsubsection{\textbf{PURSUIT~\cite{pursuit}}}
\begin{figure}[!ht]
\centering
\includegraphics[width=0.65\linewidth]{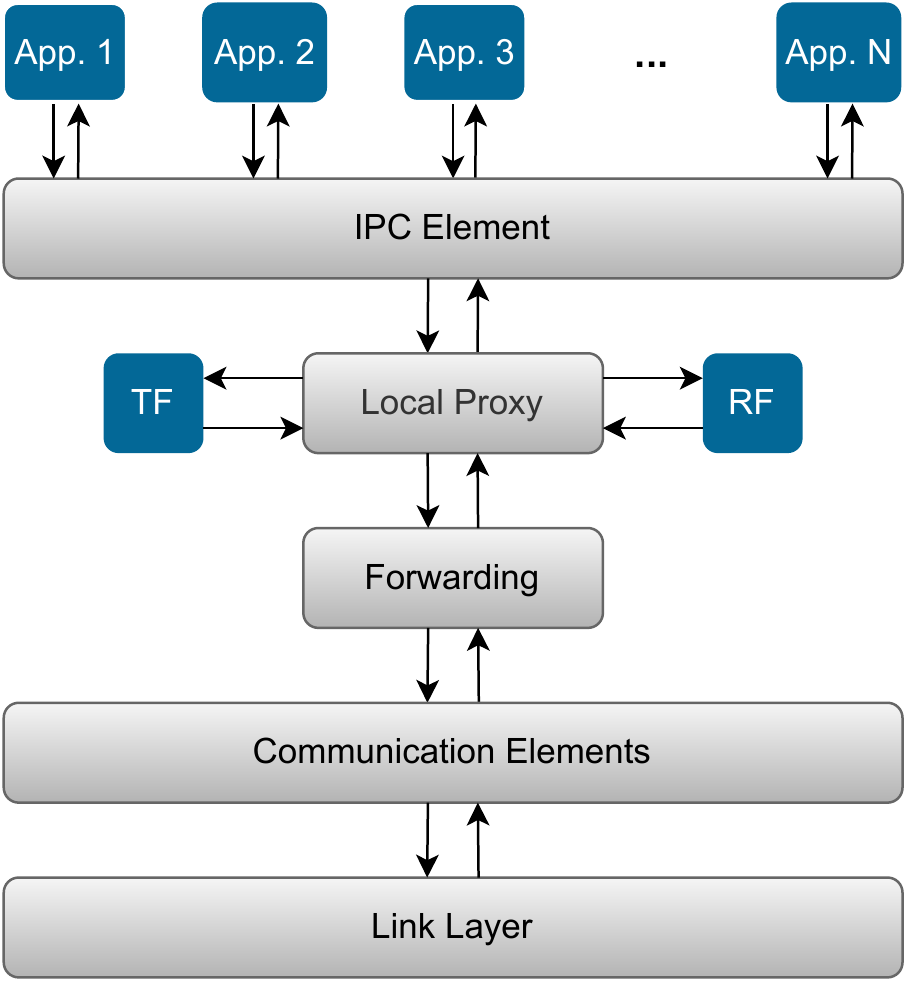}
\caption{The PURSUIT~\cite{pursuit} node structure. The core of a PURSUIT node includes three functions: Forwarding, Topology Formation and Management Function (TF) and Rendezvous Function (RF). Instead, the local proxy stores the local records.}
\label{fig:pursuit_node}
\end{figure}

is an evolution of the FP7 project \gls{PSIRP}~\cite{psirpbook} that proposes a publish-subscribe \gls{ICN} model.
Here, users interested in content subscribe to the corresponding publisher. 
A PURSUIT node encompasses several components as shown in Fig.~\ref{fig:pursuit_node}.
In the network layer, it proposes three new functions: \gls{RF}, \gls{FF} and \gls{TF}. 
The \gls{RF} is in charge of mapping subscribers to publishers and supporting name resolution.
Furthermore, the \gls{TF} collects the topology of its domain by deploying a routing protocol and exchanges information with other nodes in other domains to collect information for global routing. 
Lastly, the \gls{FF} is deployed in the \gls{FN}, which is responsible for routing the information item to the requester. 
All three functions receive dispatched requests from the local proxy, which maintains local records on the issued and pending subscriptions.
PURSUIT proposes means to identify individual information items and scoping that creates sets of individual information items and places them into a context to manipulate the information flow. 
In PURSUIT, the routers do not maintain forwarding states but use Bloom Filters for packet forwarding.
PURSUIT has been implemented as an overlay solution among multiple nodes with different locations. Therefore, it adheres to the ICN islands in IP ocean deployment scenario. 
PURSUIT ended in February 2013 as a production of the \textit{FP7} European project started in September 2010.

\subsubsection{\textbf{Network of Information (NetInf)~\cite{netinf}}}
\begin{figure}[!ht]
\centering
\includegraphics[width=\linewidth]{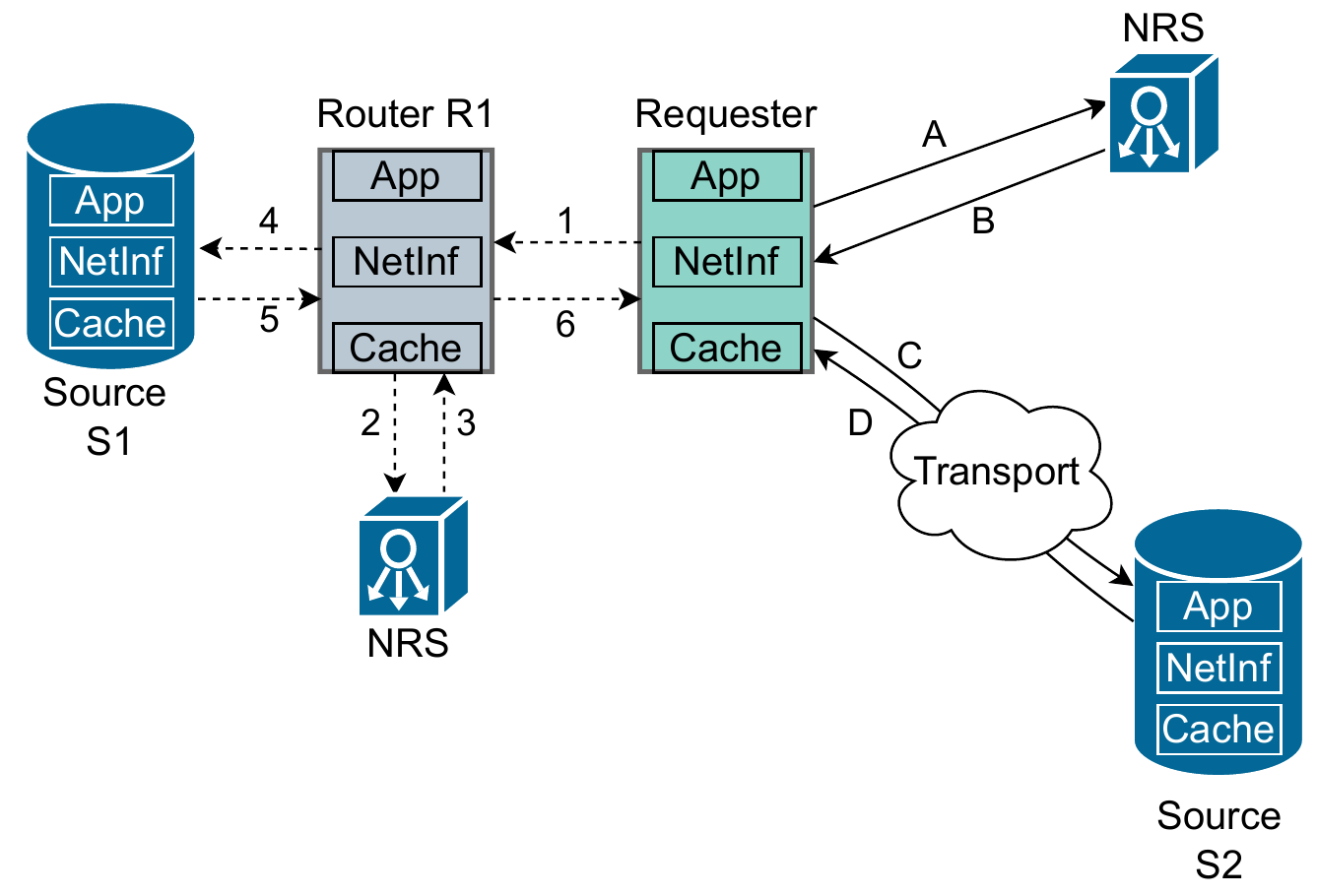}
\caption{The NetInf~\cite{netinf} architecture. NetInf introduces the Name Resolution Service (NRS) nodes that populate the routes. The numbered dot lines represent the hop-by-hop communication. Instead, the lines with the letters represent the query-based resolution for content retrieval.}
\label{fig:netinf_com}
\end{figure}

is proposed by the SAIL project~\cite{SAIL}, that introduces its naming and security model, which does not require -- but can use -- naming authorities or a \gls{PKI}. 
NetInf dedicates global \gls{NRS} nodes for name-based object retrieval.
NetInf protocol relies on \gls{NDO} and on flat named information URI schema for named information~\cite{farrell2013naming}. 
Names in NetInf contain a hash algorithm and a hash value.
NetInf adds the Convergence Layer (CL) that maps information expressed through existing protocols -- e.g., HTTP, TCP, or IP -- into specific messages compliant with a general communication paradigm. 
As shown in Fig.~\ref{fig:netinf_com}, a requester in NetInf issues name-based interest requests for content.
The request is forwarded hop-by-hop through NetInf routers until a cached copy is found or the content source is reached. 
The \gls{NRS} is queried in case of missing routes.
Alternatively (steps A-D), the requester can directly query \gls{NRS} and resolve the content name. 
The authors proposed a TCP/UDP overlay NetInf prototype. Thus, NetInf adheres to the ICN islands in IP ocean deployment scenario.
Similarly to PURSUIT, the NetInf architecture design ended in February 2013, with the end of the \textit{European FP7 project SAIL}, started in January 2010.

\subsubsection{\textbf{\acrfull{NDN}~\cite{ndn}}}
is the main \gls{ICN} project enabling content naming, routing based on names, and caching. 
In \gls{NDN}, the consumers request the contents by their names. 
An NDN router encompasses three elements---i.e., \gls{PIT}, \gls{CS}, and \gls{FIB}. 
For each request, the router checks if the requested content is present in the \gls{CS}. If this is the case, the content is directly returned to the consumer. Otherwise, the router checks the \gls{PIT} for an existing request for such content. In case an entry in \gls{PIT} exists, the router adds the interface from which such request arrived in the existing entry. Otherwise, the router forwards the request to the next node in the network by using \gls{FIB}. The routers follow the reverse path routing downstream to deliver the data to the consumer and store a copy in the \gls{CS}.
The current implementation of \gls{NDN} adheres to the overlay deployments since it is based on protocols such as CCNx, NDNLP~\cite{shi2012ndnlp} that are deployed over IP. 
The former establishes the communication over the existing \gls{UDP} transport protocol in \gls{IP} networks. Instead, the latter allows the transportation of packets between two nodes over a local one-hop link. 
Another example of the overlay implementation of \gls{NDN} is the NDN testbed~\cite{ndntestbed} which connects several \gls{NDN} nodes located in participating universities and connected through existing \gls{IP} infrastructure.
Therefore, we have considered NDN as part of ICN islands in IP ocean deployment scenario.
NDN architecture is funded by the \textit{National Science Foundation} as part of the \textit{Future Internet Architectures} program in late 2010 and is still going on. From the research perspective, NDN represents the most important and explored architecture that might bring the information-centric Internet become a real-world implementation.

\subsubsection{\textbf{NDNFlow~\cite{ndnflow}}} 
\begin{figure}[!ht]
\centering
\includegraphics[width=\linewidth]{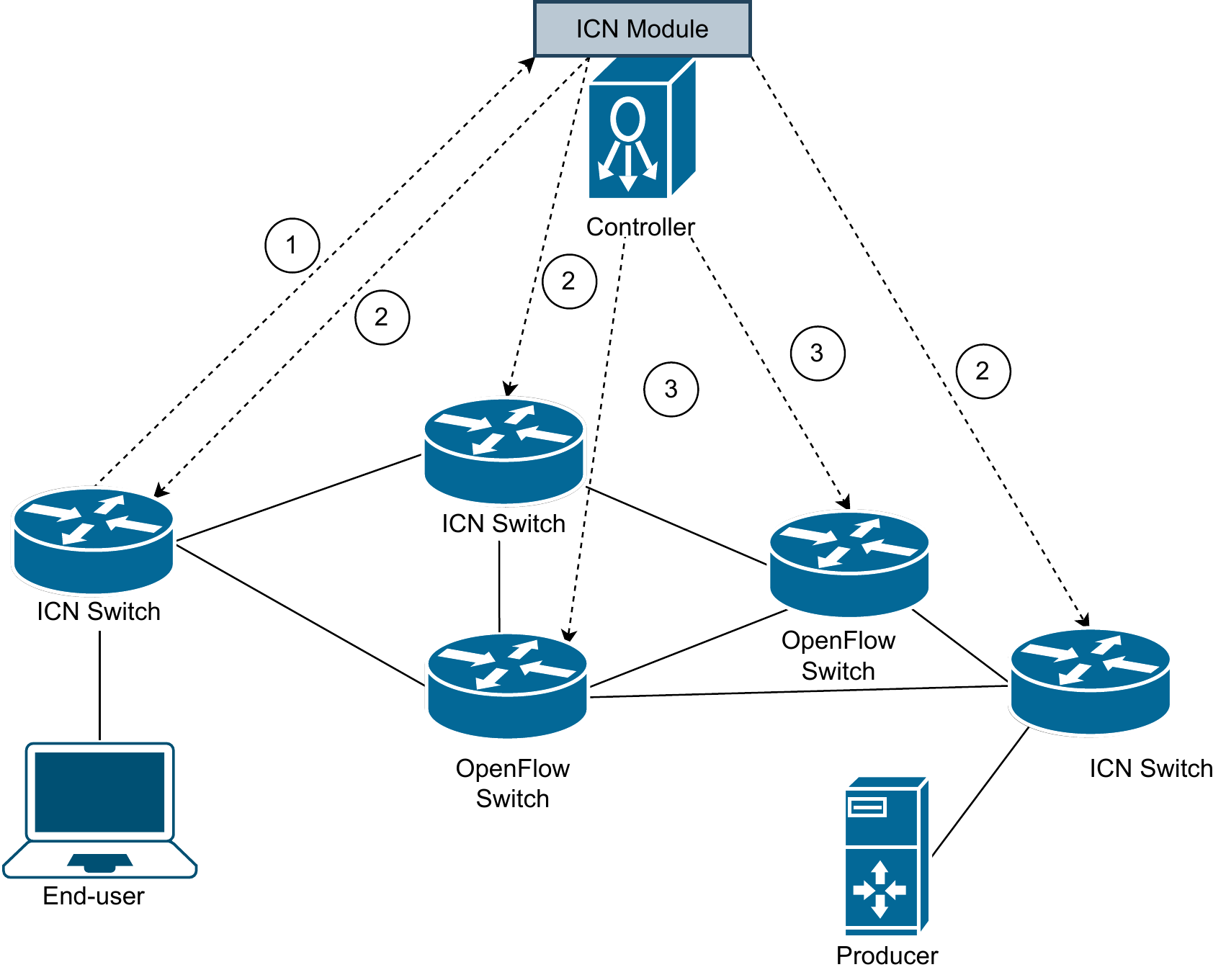}
\caption{The NDNFlow~\cite{ndnflow} architecture. Similarly to SDN, NDNFlow charges a controller for managind and controlling the communication rules forboth ICN and OpenFlow switches on the data plane.}
\label{fig:ndnflow_com}
\end{figure}

is a coexistence architecture inspired by \gls{SDN} paradigm. 
It introduces a new application-specific layer to the OpenFlow switch that can handle the \gls{ICN} traffic separately from the standard \gls{IP} traffic. 
Furthermore, the controller has an ICN module to process and compute paths for ICN traffic. The network in NDNFlow includes legacy SDN and ICN-aware switches, where both types communicate with the controller through two dedicated communication channels. The unreachable ICN-aware switches establish communication through IP-encapsulated tunnels, and the OpenFlow controller configures the IP switches to properly forward these tunnels.
In NDNFlow architecture, there are both ICN-aware switches and legacy OpenFlow switches as represented in Fig.~\ref{fig:ndnflow_com}. Upon reception of a request, the ICN-aware switch forwards it to the ICN module of the controller (Step 1). 
The ICN module uses the network topology information and ICN functionality to compute the ICN flow paths. Eventually, it properly configures the added routes in the ICN-aware switches (Step 2). Lastly, the ICN module instructs the OpenFlow controller to configure the IP rules on the other intermediate legacy OpenFlow switches (Step 3). 
NDNFlow has been implemented in a testbed environment where different experiments have been conducted to evaluate this architecture.
Lastly, we have considered NDNFlow as part of the ICN islands in IP ocean deployment scenario.
NDNFlow architecture is designed in 2015 and the source code\footnote{ \url{https://github.com/TUDelftNAS/SDN-NDNFlow}} is released for future research.

\subsubsection{\textbf{Overlay for Information Centric Networking (O-ICN)~\cite{oicn}}}
\begin{figure}[!ht]
\centering
\includegraphics[width=0.9\linewidth]{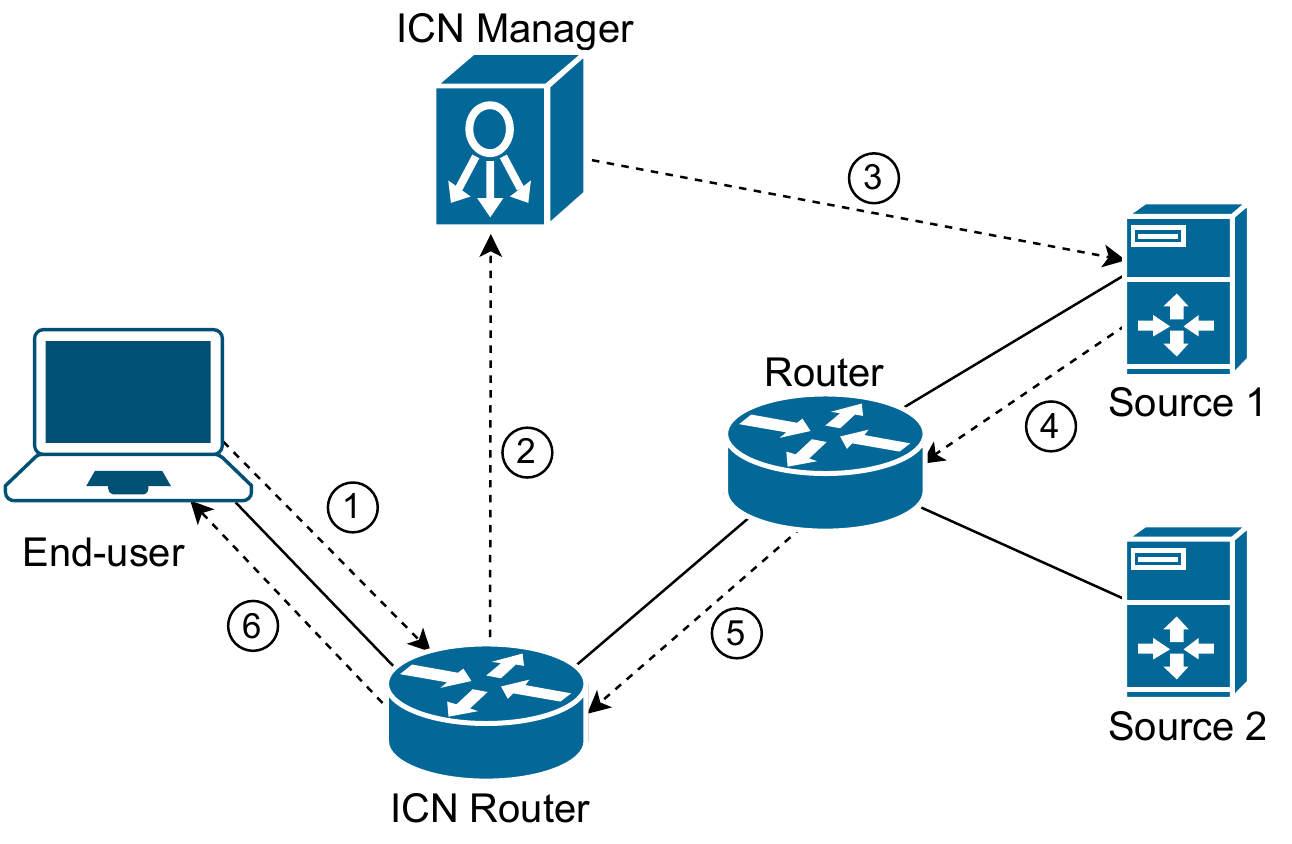}
\caption{The O-ICN~\cite{oicn} architecture. Its design includes an ICN controller -- ICN Manager -- that indicates rules to the ICN routers, without introducing changes for the IP routers.}
\label{fig:oicn_com}
\end{figure}

is an overlay architecture that makes use of \gls{SDN} concept, and the core component of its deployment is the ICN Manager, similar to the controller in \gls{SDN}.
Aside from ICN Manager, the other communicating parties in an O-ICN network include end-users, sources, \gls{IP} and \gls{ICN} routers.
The ICN Manager is an extended version of the \gls{DNS} server, and as such, it provides name resolution services for the \gls{ICN} requests and \gls{DNS} resolution by resolving the requested IP address for non-ICN requests. 
O-ICN adds a layer between the Application and Transport layer of the TCP/IP protocol stack, called \gls{ICN} sublayer that accommodates \gls{ICN} communication exchange by modifying some fields on \gls{IP} and \gls{DNS} packets.
The \gls{ICN} routers inspect the packets beyond the \gls{IP} layer -- i.e., \gls{ICN} sublayer -- and caches the content. 
For caching decisions and changes on the stored content, the router contacts the \gls{ICN} Manager.
%
The communication flow in O-ICN is represented in Fig.~\ref{fig:oicn_com}. For each \gls{ICN} request made by the end-user, the edge router contacts the \gls{ICN} Manager, which resolves the content's source and communicates the user's address. 
Then, the source is in charge of routing the requested content to the user. For \gls{IP} requests, existing \gls{DNS} resolve content's location, and afterward, existing TCP/IP routing mechanisms are followed. 
There are two coexistence scenarios of \gls{ICN} and \gls{SDN}: i) the \gls{ICN} Manager and \gls{SDN} controllers function separately and ii) the functionalities of the \gls{ICN} Manager and \gls{SDN} controller are merged.
In~\cite{agrawal2018icn} the O-ICN is implemented and evaluated as an ns-3-based simulator -- i.e., OICNSIM -- where each O-ICN component is represented as a helper class. Afterward, the authors studied its performance under different ICN caching policies.
Thus, O-ICN is considered to adhere to the ICN islands in IP ocean deployment scenario.
From the research perspective, the open-source OCINSIM\footnote{ \url{https://github.com/TCS-Research/OICNSIM}} simulator proposed during the design of O-CIN offers the possibility for the researchers to explore the coexistence issue.

\subsubsection{\textbf{CONET~\cite{conet}}}
\begin{figure}[!ht]
\centering
\includegraphics[width=\linewidth]{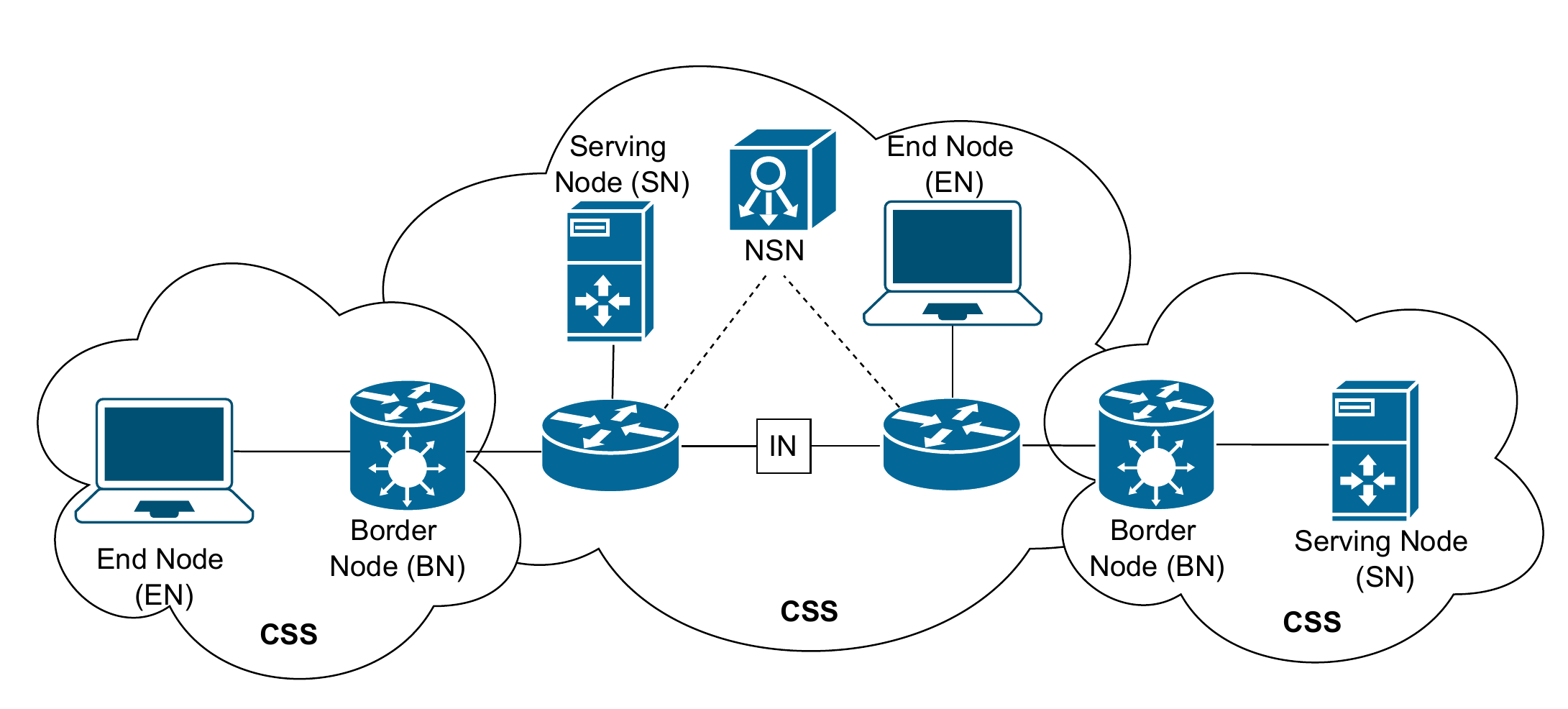}
\caption{The architecture overview for CONET~\cite{conet}. It is composed of different CONET Sub Systems (CSS) interconnected using Border Nodes (BN).}
\label{fig:conet}
\end{figure}

is an information-centric architecture that provides users with network access to remote named resources -- i.e., named data or service access points -- and interconnects CONET Sub Systems (CSS). 
Nodes in CONET are classified as \gls{NSN}, Serving Nodes (SN), Border Nodes (BN), Internal Nodes (IN), and End Nodes (EN), as represented in Fig.~\ref{fig:conet}. Each CONET node has the protocol stack containing CONET and Under-CONET layers. 
CONET splits the content into different chunks and inserts them into a named data CONET Information Units (CIU).
The EN retrieves named data from SN through---i.e., EN issue interest CIU and SN respond with named-data CIU.
BN is located at the border between CSS that forwards the carrier packets and caches named data CIU. 
IN is placed chiefly inside a CSS network and provides in-network caches. Finally, \gls{NSN} enable the CONET routing-by-name process.
The EN requests named data, and the intermediate BN resolves the CSS address -- an Under CONET address, e.g., IPv4 address -- of the next BN closest to the SN of the other CSS.
Whenever the packet reaches the CSS network, the IN is in charge of parsing the packet and forwarding it using the under-CONET engine. Then, the content is retrieved by the first node that can provide it without further propagation in the CSS network. The content is delivered to the requester following the reverse path.
CONET can be implemented as \gls{ICN} clean slate approach, overlay, or hybrid. In the latter, both IN and BN of the IP CSS can be modified to accommodate the CONET traffic. In particular, these nodes have the fast forwarding path, which handles the forwarding of CONET and \gls{IP} traffic. They also have routing tables containing IP net prefixes and CONET named prefixes. 
Given the above considerations of the CONET design, we consider it as part of the ICN islands in IP ocean and ICN, and IP islands in IP ocean deployment scenario.
Lastly, since the CONET's source code is freely released\footnote{ \url{https://github.com/StefanoSalsano/alien-ofelia-conet-ccnx}}, it can be used in the future to conduct more research on the overlay coexistence architectures.

\subsubsection{\textbf{GreenICN~\cite{greenICN}}}  
\begin{figure}[!ht]
\centering
\includegraphics[width=0.9\linewidth]{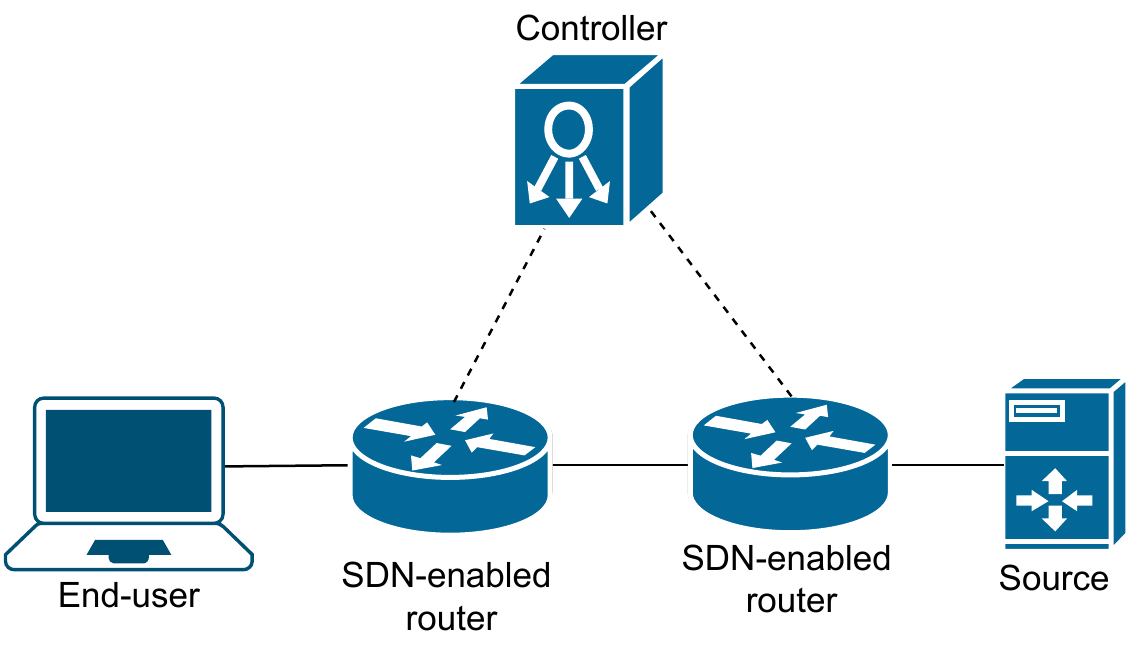}
\caption{The GreenICN~\cite{greenICN} architecture. It inherits SDN features such as the controller and data plane switches, which are enabled to parse ICN packets. The controller assists in case of missing rules.}
\label{fig:greenicn}
\end{figure}

is an \gls{SDN}-based \gls{ICN} architecture, and similarly to O-ICN, it distinguishes between the control and the data plane. 
The SDN-enabled routers are modified to accommodate ICN packets to forward the \gls{ICN} requests. 
GreenICN introduces four elements for such accommodation: an ICN protocol identifier contained in the ICN packets, a publicly routable network address per domain, the object's name, and a MsgID. This object name determines the routing path while the MsgID replaces the destination IP in the requests and the source IP in the responses.
A consumer requests content issuing an IP packet where the routable public address is set as the destination IP and the destination port is set to ICN protocol identifier. 
Then, the SDN-enabled router checks if the packet matches one of the saved rules. 
Otherwise, the packet is forwarded to the controller that analyses the ICN payload and extracts the requested object name from it. 
Then, the controller changes the IP and port number of the packet to the cache's IP for the destination IP address and MsgID for the source IP address. Finally, the controller installs the forwarding rules on the elements on the path to the content and sends the packet back to the SDN-enabled router. 
Downstream, the response packet is forwarded back to the requester by performing the address and port rewrite on the egress SDN node. At the end of the response delivery to the requester, the SDN node notifies the controller that all states of MsgID should be removed. Fig.~\ref{fig:greenicn} presents a simplified view of GreenICN architecture. 
GreenICN has been implemented on Terma OpenFlow controller-framework~\cite{trema} and CCNx library. Thus, it adheres to the ICN islands in IP ocean and ICN, and IP islands in IP ocean deployment scenarios.
Lastly, GreenICN design started in 2013 and ended in 2013 with the EU project on \textit{GreenICN: Architecture and Applications of Green Information Centric Networking}.

\subsubsection{\textbf{coCONET~\cite{coCONET}}} 
was inspired by CONET architecture~\cite{conet} and SDN. 
Like SDN, coCONET's design decouples the data plane, composed of ICN nodes -- i.e., OpenFlow switches -- and end-users, from the control plane, consisting of \gls{NRS} nodes---i.e., OpenFlow Controller.
coCONET presents an extension of the OpenFlow protocol to ensure the forwarding by name, security, and proactive caching.
The communication in coCONET generally involves both \gls{ICN} and \gls{NRS} nodes. An \gls{ICN} node contacts an \gls{NRS} node when an event is triggered---e.g., an interest or a content packet arrives. 
Conversely, the NRS node contacts the \gls{ICN} node for logic operations based on a timeout. 
Here, the end-users issue an interest request that reaches an \gls{ICN} node. 
In case of missing routes, the \gls{ICN} node queries the \gls{NRS} nodes. 
coCONET enforces the ICN nodes to verify the content integrity and authenticity before forwarding it to the requester.
coCONET has been implemented using SDN switches and controllers software, and in the future, it will be implemented in the OFELIA testbed~\cite{kopsel2011ofelia}.
Therefore, we consider it as part of the ICN islands in IP ocean and ICN, and IP islands in IP ocean deployment scenario.
Lastly, coCONET was proposed in 2012 under the \textit{EU Project Convergence} which ended in 2013.
\subsubsection{\textbf{OpenFlow-ICN~\cite{openflowicn}}}  
\begin{figure}[!ht]
\centering
\includegraphics[width=0.9\linewidth]{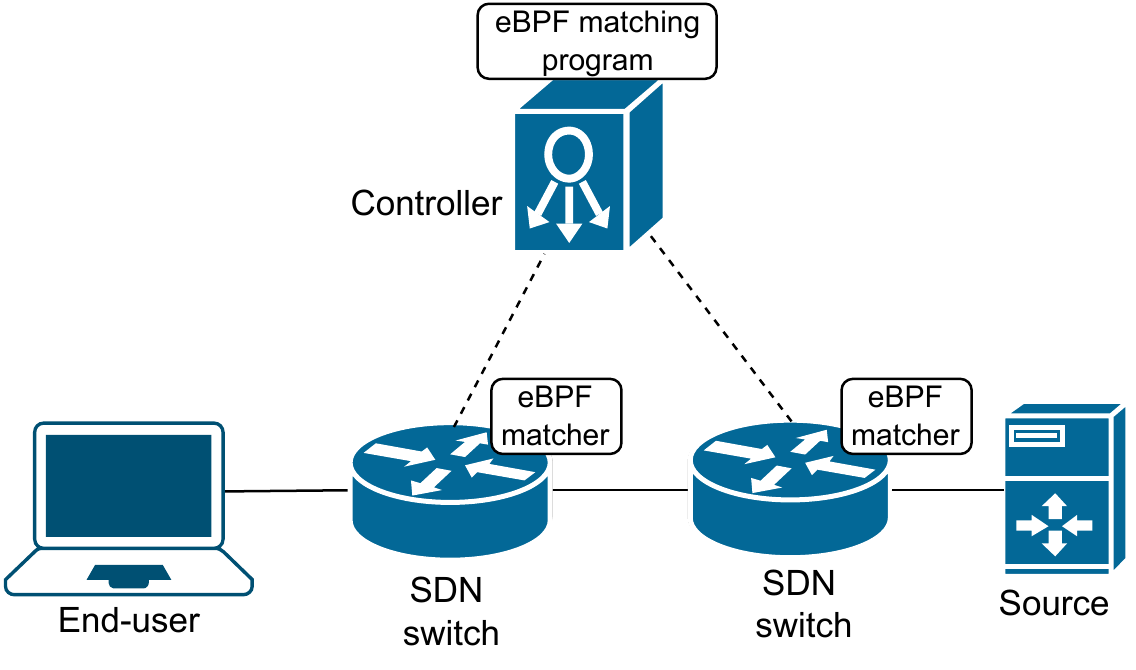}
\caption{The OpenFlow-ICN~\cite{openflowicn} architecture. Its controller encompasses the Extended Berkeley Packet Filters (eBPF) module to accommodate the ICN semantics in IP packets.}
\label{fig:openflowicn}
\end{figure}

leverages SDN concepts, and in particular, the OpenFlow protocol.
OpenFlow-ICN extends the OpenFlow protocol and uses the Extended Berkeley Packet Filters (eBPF)~\cite{mccanne1993bsd} to accommodate ICN packets in IP-based networks. Generally, matching on IP packets is checked on specific fields -- e.g., destination IP address or port number -- while here, the match is performed on the value turned after the execution of the eBPF program. In this context, OpenFlow-ICN modifies the SDN controllers to recognize ICN packets and deal with complex and large eBPF matching programs. Fig.~\ref{fig:openflowicn} presents OpenFlow-ICN architecture. 
For the forwarding plane, OpenFlow-ICN adopts a multi-switch forwarding controller app for Ryu~\cite{van2016backup}, and it handles the IP traffic. 
On the other side, the ICN app handles routing for ICN traffic by mapping name prefixes to a particular host. 
For an interest request, the switch checks for prefix matching.
In case of missing configurations, it contacts the controller that finds the most suitable location to set up a route. Whenever a node performs a local multicast of the packet, the switch replaces all multicast MAC and IP addresses with the address of the next ICN hop on the path. For this purpose, a static table containing the locations and addresses of the local routers is maintained. Therefore, in OpenFlow-ICN, short tunnels are created to forward the interests to the producer of the requested data.
OpenFlow-ICN adds the MAC and IP forwarding rules to the intermediate switches. In this way, the data packets follow the reverse path routing.
All intermediate nodes can cache data in the path toward the requester. 
For evaluation purposes, OpenFlow-ICN has been implemented on SciNet testbed connected to TNO NDN testbed~\cite{fan2015managing}. Lastly, we consider OpenFlow-ICN part of the ICN islands in IP ocean and ICN, and IP islands in IP ocean deployment scenario.

\subsubsection{\textbf{SDN-PURSUIT~\cite{SDNPURSUIT}}}
extends the PURSUIT proposal~\cite{pursuit} and exploits the SDN concept. 
It uses the same network components as PURSUIT -- i.e., \gls{RF}, \gls{TF} and \gls{FF} and it slightly changes the latter.
In addition, SDN-PURSUIT inherits from PURSUIT the information aggregation through scoping. 
The \gls{RF} matches the publishers with the subscribers, manages the information graph, and initiates the creation of a forwarding path by sending a request to the \gls{TM}, which in turn creates such path expressed through LIPSIN identifiers~\cite{jokela2009lipsin}. 
The presence of the \gls{SDN} concept requires the need of the flow notion in the \gls{ICN} context, and this is the novelty that SDN-PURSUIT proposes compared to PURSUIT. 
Here, it uses a forwarding scheme based on some identifiers rather than directly on content names.
To remove the source and destination IP addresses, in SDN-PURSUIT, all nodes in the network have to be configured into promiscuous mode to capture all flowing traffic. However, when a packet arrives in a specific port, it is undoubtedly destined for that node since it is the controller that maps the local ports to the datapaths installed on the switches.
For the implementation of SDN-PURSUIT, NITOS~\cite{pechlivanidou2014nitos} and OFELIA~\cite{kopsel2011ofelia} testbeds are used.
SDN-PURSUIT adheres to ICN islands in IP ocean deployment scenario.
Similarly to PURSUIT, SDNPURSUIT was proposed under the \textit{EU FP7 project PURSUIT}, which ended in 2013.
\subsection{SP analysis}
\label{ssec:overlay-sp-analysis}
\subsubsection{\textbf{Trust}}
Similarly to publish-subscribe architectures~\cite{fiege2004security, esposito2014security}, trust in \textbf{PURSUIT} is not based on securing producers or consumers, but it is left to the application instead. 
On the other hand, PURSUIT uses flat names that are self-certifying and can be used to establish trust based on content, allowing verification procedures based on content. Additionally, PURSUIT design includes scoping field and dissemination strategies to delimit the group of producers and consumers on the application level and control the dissemination of notifications within the infrastructure, creating the so-called group of trustees.
In \textbf{NetInf}, trust is based directly on the content, and during the design, the authors considered providing content-based security.
Here, the NDO contains security-related information -- e.g., NDO hash value or producer's public key hash -- ensuring data integrity.
By including the hash of the producer's public key, NetInf enables the producer's pseudonymity. Each NDO producer can establish trust based on the pseudonym, and all the content consumers trust the content published under this pseudonym. Optionally, the architecture provides the owner identification feature, revealing the real-world producer's identity by binding it to its pseudonym. However, NetInf core architecture does not provide this feature directly. But instead, it relays in standard \gls{PKI}---e.g., Trusted Third Party systems or Web of Trust~\cite{caronni2000walking}.
Additionally, the CL is a fundamental component of NetInf, and its correct operation provides reliability to NetInf architecture since it maps different protocol abstractions to a single unique abstraction used by NetInf while maintaining all security properties of the data.
\textbf{NDN} adheres to the data-centric security model where the \gls{SP} procedures are guaranteed based on content itself~\cite{zhang2018overview}. In the NDN's overlay implementation, the \gls{NDN} islands establish trust in content.
Here, NDN's trust model resembles the Simple Distributed Security Infrastructure (SDSI/SPKI)~\cite{rivest1996simple} in trust anchor establishment.
Furthermore, NDN enables the application to define naming conventions to systematically construct the names of the cryptographic keys or certificates used for signing, verification, encryption, and decryption.
Instead, the current \gls{IP} ocean is used for communication between such islands. Thus existing infrastructures can be used to establish trust in the communication between islands. 
The described trust model for NDN islands can also be applied for \textbf{NDNFlow} architecture.
However, due to the centralized nature of \textbf{NDNFlow}, the controller is critical to ensuring trust. It coordinates and supervises the information flow between communication parties. Additionally, the \gls{ICN} modules at the controller can enable trust for ICN traffic by securing a single content object or even the entire namespace by embedding security features directly into content names. 
Similarly, in \textbf{O-ICN}, \textbf{GreenICN}, \textbf{OpenFlow-ICN} and \textbf{SDN-PURSUIT} trust is mainly based on the controller---i.e., alternatively called ICN Manager in O-ICN. Generally, the controller ensures name resolution, link management, and routing. In particular, for GreenICN, the controller comprises different parts, including switch daemon, path manager, packet filter, and topology modules. 
Regarding \textbf{CONET}, its overlay implementation ensures trust mainly by securing the communication channel between islands through PKI mechanisms. Instead, in the integration approach, the trust model for \gls{ICN} CSS changes and follows the content-based model described for NDN islands. 
In the \textbf{coCONET} design, content-based security has been discussed and motivated. coCONET discusses the use of a PKI infrastructure for managing the key distribution. Furthermore, the NRS node manages the association between names and public keys whenever human-readable names are used.

\subsubsection{\textbf{Data origin authentication}}
In \textbf{PURSUIT's} ICN islands, publishers sign each content packet. 
Here, a consumer retrieves the certificate of its producer, which in turn points to its signer's certificate and finally reaches the trust anchor. The origin is validated if all the certificates in this chain are valid and comply with the trust policies.
The same rationale follows for \textbf{NetInf}, \textbf{NDN}, \textbf{NDNFlow}, \textbf{O-ICN}, \textbf{OpenFlow-ICN} and \textbf{SDN-PURSUIT}.
These architectures must establish the communication between different \gls{ICN} islands in tunnel mode to ensure the data origin authentication. However, none properly design the authentication for the origin of data flowing in different islands that use various PKI mechanisms.
Conversely, the deployment scenarios for \textbf{GreenICN}, \textbf{CONET} and \textbf{coCONET} include not only \gls{ICN} islands, but also \gls{IP} islands. Here, the complexity of data origin authentication further increases. coCONET tries to overcome this issue and demands each ICN node to verify the signatures before further forwarding the content. However, this solution is not lightweight since it requires all ICN nodes to obtain the public keys associated with all requested contents.

\subsubsection{\textbf{Peer entity authentication}}
%
All the analysed overlay architectures include ICN islands in their deployment scenarios. Therefore, the lack of peer entity authentication is inherited from ICN design in these architectures.
%
An exception make the design of \textbf{GreenICN}, \textbf{CONET} \textbf{coCONET} that include also IP islands. 
Like native \gls{IP} networks, in \gls{IP} CSS parts, CONET and coCONET can establish peer entity authentication through existing mechanisms -- e.g., certificates -- that bind hosts' identity to their corresponding public key.

\subsubsection{\textbf{Data integrity}}
%
Like data origin authentication, data integrity for all the considered overlay architectures in the \gls{ICN} islands is based on the producer's digital signatures. 
In particular, \textbf{NetInf} adds the hash value of the transmitted NDO, ensuring data integrity. Instead, the hash of the producer's public key is included in the packet for the dynamic content---i.e., content that might change in time. Eventually, the consumer can calculate the hash of the public key, compare it with the received one, and then verify the data integrity using the producer's public key.
For the IP islands present in \textbf{GreenICN}, \textbf{CONET} and \textbf{coCONET}, IPSec solutions -- e.g., HMAC -- can be used to ensure such feature. 

\subsubsection{\textbf{Authorization \& access control}}
This feature cannot be ensured in all the \gls{ICN} islands of the analysed overlay architectures. 
In the \textbf{NetInf} design, the authors confirmed that the proposed architecture could not apply access control over levels of information, leaving open access without any restriction.
%
For \textbf{NDNFlow}, \textbf{O-ICN}, \textbf{GreenICN}, \textbf{coCONET}, \textbf{OpenFlow-ICN} and \textbf{SDN-PURSUIT}, the lack of authorization \& access control policies can be attained by taking advantage of the controller.
In particular, such policies are primarily enforced on flow-level information by the controller~\cite{SDNSecurity}. Other solutions consider the controller or the middle-boxes -- i.e., firewalls -- to implement such policies. 
Different access control systems for OpenFlow networks have been proposed---i.e., Ethane~\cite{casado2009rethinking}, Resonance~\cite{nayak2009resonance}.
However, modifications must be made to accommodate the \gls{ICN} semantics and to achieve complete access control for both types of traffic---i.e., \gls{IP} or \gls{ICN}.
Lastly, \textbf{CONET} can construct \gls{ACL} in the border nodes.  

\subsubsection{\textbf{Accountability}}
%
All the overlay architectures that include \gls{ICN} islands in their deployment ensure accountability only for producers since they sign each data packet. Conversely, they can not achieve accountability for consumers.
Nevertheless, some of the analysed architectures can exploit different elements included in their design to ensure such features. 
For example, \textbf{PURSUIT} might take advantage of \gls{RP} to keep track of the network's activities.
Similarly, the controller in \textbf{NDNFlow}, \textbf{GreenICN}, \textbf{O-ICN}, \textbf{OpenFlow-ICN} and \textbf{SDN-PURSUIT} could be used to trace the actions taken by OpenFlow switches and also all the hosts it observes. 
On the other hand, the presence of known attacks -- i.e., \gls{IP} or \gls{MAC} spoofing -- the achievement of this \gls{SP} feature can be compromised for the between-islands communication and the IP islands of \textbf{GreenICN}, \textbf{CONET} and \textbf{coCONET}.

\subsubsection{\textbf{Data confidentiality}}
None of the proposed overlay architectures considers the key management and appropriate encryption schemes that can be used to achieve data confidentiality. 
Since all the analysed architectures encompass ICN islands, the basic approach to data confidentiality is encryption.
In IP islands, the Diffie-Hellman key exchange protocol~\cite{li2010research} is used to derive encryption keys for point-to-point sessions.
Nevertheless, Diffie-Hellman does not apply to construct encryption keys for multi-party communications---e.g., NDN applications~\cite{mosko2017mobile}. 
Here, structured names can be used to create new encryption schemes -- e.g., attribute-based encryption (NAC-ABE)~\cite{zhang2018nac} -- which allow the key distribution for multi-party communication applications.
Most of the analysed architectures design include using human-readable names in the ICN island, introducing possible information leakage~\cite{bardhi2021icn}.
Additionally, traffic analysis attacks might be present in IP and ICN islands.
Instead, SDN-based overlay architectures -- i.e., \textbf{NDNFlow}, \textbf{GreenICN}, \textbf{O-ICN}, \textbf{OpenFlow-ICN} and \textbf{SDN-PURSUIT} --
can exploit the controller for managing and monitoring the encryption schemes.

\subsubsection{\textbf{Availability}}
The \gls{ICN} islands present in all the overlay architectures inherit well-known \gls{ICN} availability issues---e.g., flooding attacks, DoS and DDoS, timing attacks, content poisoning, and cache pollution~\cite{ambrosin2018security}.
Additionally, new elements might open the door for new availability issues. 
For example, if the RP in \textbf{PURSUIT} gets compromised, it exposes all the communication flow. Also, an attacker can perform a \gls{DoS} attack by modifying or deleting data in the local proxy tables. 
Another example in \textbf{NetInf} is \gls{NRS}, where a compromised NRS in one NetInf domain would make the request flow to other local \gls{NRS} or the global one, compromising the network performance.
Instead, the controller in SDN-based architectures -- i.e., \textbf{NDNFlow}, \textbf{GreenICN}, \textbf{O-ICN}, \textbf{OpenFlow-ICN} and \textbf{SDN-PURSUIT} -- represents a single point of failure. Here, the attacker exploits the presence of too many rules indicated by the controller and limited storage capacities to follow such rules~\cite{alsmadi2015security, SDNSecurityProsCons}.
Furthermore, the attacker can tempt to flood the communication channel between the controller and switches to interrupt benign network activities. 
Lastly, an exception makes \textbf{coCONET} design that requires each ICN node to verify the signature before forwarding the content. 
Although time-consuming and computationally expensive, this requirement helps prevent \gls{DoS} and \gls{DDoS} attacks.

\subsubsection{\textbf{Anonymous communication}}
Most of the analysed architectures inherit the consumer's anonymity from \gls{ICN}, while the producer's signature can reveal its identity.
Additionally, \textbf{NetInf} also provides owner pseudonymity to ensure such anonymity for producers.
In \textbf{CONET}, attackers might use the BN and IN nodes to deanonymise the communicating parties since they contain information related to nodes participating in the communication.

\subsubsection{\textbf{Traffic flow confidentiality}}
IP and ICN islands in the analysed overlay architectures suffer traffic analysis and side-channel attacks that aim to find statistical patterns in the exchanged traffic, violating the confidentiality of traffic flow.
%
It is shown that the control plane of SDN-based architectures might leak information since the encryption scheme in OpenFlow is optional.
Here, lousy usage -- or even lack -- of encryption schemes could open the door for possible Man-in-the-the-Middle (MiM) attacks~\cite{benton2013openflow}. 
Furthermore, name-based forwarding and hierarchical, human-readable names generally expose the users to privacy threats~\cite{bardhi2021icn}.

\subsection{Comparison}
\label{ssec:overlay-comparison}
Table~\ref{tab:overlay-evaluation} reports the evaluation scores for the analysed overlay architectures for each of the ten SP features.  
%
%
Among them, \textbf{coCONET} ensures highest scores. 
Indeed, during the coCONET design, the authors made multiple security and privacy considerations, as we have extensively described during the SP analysis for such architecture. 
In particular, they motivate and describe the mechanisms that enable data origin authentication and integrity features.
Nevertheless, as with most of the overlay architectures, also coCONET fails to achieve data and traffic flow confidentiality and anonymous communication. 
On the other hand, the architectures that are proposed as new information-centric architectures that additionally discuss their coexistence with IP protocol -- i.e., \textbf{PURSUIT}, \textbf{NDN}, \textbf{CONET}, except for \textbf{NetInf} -- reach the lowest average score.
Indeed, such architectures are designed without considering the SP implications that the coexistence of ICN and IP protocols might bring.
Conversely, the SDN-based architectures -- i.e., \textbf{NDNFlow}, \textbf{SDN-PURSUIT}, \textbf{O-ICN}, \textbf{GreenICN} and \textbf{OpenFlow-ICN} -- score slightly higher scores than the previously cited architectures since they leverage additional components -- e.g. controller -- to establish SP procedures. 
Instead, in terms of feature coverage, \textbf{trust}, \textbf{data origin authentication and integrity}, and \textbf{accountability} are amongst the ensured features by at least partially by almost all the analysed architectures.
Overall, according to our analysis, the majority of the considered overlay architectures fail to ensure \textbf{peer entity authentication}, \textbf{authorization and access control}, \textbf{data confidentiality}, \textbf{availability}, \textbf{anonymous communication} and \textbf{traffic flow confidentiality}.
\begin{table}[!htb] 
\centering
\caption{Evaluation of security and privacy features of overlay IP-ICN coexistence architectures.} 
\label{tab:overlay-evaluation}
\resizebox{\columnwidth}{!}{
    \begin{tabular}{c | c c c c c c c c c c}
      \toprule
      \textbf{\makecell{SP Feature}} & 
      \textbf{\makecell{\rotatebox{90}{\scriptsize PURSUIT~\cite{pursuit}}}} & 
      \textbf{\makecell{\rotatebox{90}{\scriptsize NetInf~\cite{netinf}}}} & 
      \textbf{\makecell{\rotatebox{90}{\scriptsize NDN~\cite{ndn}}}} &
      \textbf{\makecell{\rotatebox{90}{\scriptsize NDNFlow~\cite{ndnflow}}}} &
      \textbf{\makecell{\rotatebox{90}{\scriptsize O-ICN~\cite{oicn}}}} &
      \textbf{\makecell{\rotatebox{90}{\scriptsize CONET~\cite{conet}}}} &
      \textbf{\makecell{\rotatebox{90}{\scriptsize GreenICN~\cite{greenICN}}}} &
      \textbf{ \makecell{\rotatebox{90}{\scriptsize coCONET~\cite{coCONET}}}} &
      \textbf{\makecell{\rotatebox{90}{\scriptsize OpenFlow-ICN~\cite{openflowicn}}}} &
      \textbf{\makecell{\rotatebox{90}{\scriptsize SDN-PURSUIT~\cite{SDNPURSUIT}}}} \\
      \toprule 
       \makecell{Trust} & \tiny \pie{180} & \tiny \pie{360} & \tiny \pie{180} & \tiny \pie{180} & \tiny \pie{180} & \tiny \pie{180} & \tiny \pie{180} & \tiny \pie{360} & \tiny \pie{180} & \tiny \pie{180} \\ 
        \midrule
        \makecell{\thead{Data origin \\ authentication}} & \tiny \pie{180} & \tiny \pie{180} & \tiny \pie{180} & \tiny \pie{180} & \tiny \pie{180} & \tiny \pie{180} & \tiny \pie{180} & \tiny \pie{360} & \tiny \pie{180} & \tiny \pie{180} \\ 
        \midrule
        \makecell{\thead{Peer entity \\ authentication}}& \tiny \pie{0} & \tiny \pie{0} & \tiny \pie{0} & \tiny \pie{0} & \tiny \pie{0}  & \tiny \pie{180} & \tiny \pie{180} & \tiny \pie{180} & \tiny \pie{0} & \tiny \pie{0} \\ 
        \midrule
        \makecell{Data integrity} & \tiny \pie{180} & \tiny \pie{180} & \tiny \pie{180} & \tiny \pie{180} & \tiny \pie{180} & \tiny \pie{180} & \tiny \pie{180} & \tiny \pie{360} & \tiny \pie{180} & \tiny \pie{180} \\ 
        \midrule
        \makecell{\thead{Authorization \\ \& access control}} & \tiny \pie{0} & \tiny \pie{0} & \tiny \pie{0} & \tiny \pie{180} & \tiny \pie{180} & \tiny \pie{0} & \tiny \pie{180} & \tiny \pie{180} & \tiny \pie{180} & \tiny \pie{180} \\ 
        \midrule
        \makecell{Accountability} & \tiny \pie{180} & \tiny \pie{180} & \tiny \pie{180} & \tiny \pie{180} & \tiny \pie{180} & \tiny \pie{180} & \tiny \pie{180} & \tiny \pie{180} & \tiny \pie{180} & \tiny \pie{180} \\ 
        \midrule
        \makecell{\thead{Data \\ confidentiality}} & \tiny \pie{0} & \tiny \pie{0} &  \tiny \pie{0} & \tiny \pie{0} & \tiny \pie{0}  & \tiny \pie{0} & \tiny \pie{0} & \tiny \pie{0} & \tiny \pie{0} & \tiny \pie{0} \\ 
        \midrule
        \makecell{Availability} & \tiny \pie{0} & \tiny \pie{0} & \tiny \pie{0} & \tiny \pie{0} & \tiny \pie{0} & \tiny \pie{0} & \tiny \pie{0} & \tiny \pie{180} & \tiny \pie{0} & \tiny \pie{0} \\ 
        \midrule
        \makecell{\thead{Anonymous \\ communication}} & \tiny \pie{180} & \tiny \pie{360} & \tiny \pie{180} & \tiny \pie{180} & \tiny \pie{180} & \tiny \pie{0} & \tiny \pie{0} & \tiny \pie{0} & \tiny \pie{180} & \tiny \pie{0} \\ 
        \midrule
        \makecell{\thead{Traffic flow \\ confidentiality}} & \tiny \pie{0} & \tiny \pie{0} & \tiny \pie{0} & \tiny \pie{0} & \tiny \pie{0}  & \tiny \pie{0} & \tiny \pie{0} & \tiny \pie{0} & \tiny \pie{0} & \tiny \pie{0} \\ 
    \bottomrule
    
    \end{tabular}}
\\[1\baselineskip]
\tiny \pie{360} \scriptsize Fulfilled~~~~~~ \tiny \pie{180} \scriptsize{Partially fulfilled}~~~~~~\tiny \pie{0} \scriptsize{Not fulfilled}
\end{table}

\section{Underlay Architectures}  
\label{sec:underlay}
%
In this section, we present the analysis of the coexistence architectures adhering to the underlay deployment approach. Such architectures introduce new components -- i.e., proxies or gateways -- nearby the \gls{IP} and \gls{ICN} islands to ensure their coexistence. 
We first describe the underlay architectures (Section~\ref{ssec:underlay-description}). Then, we present the SP analysis of ten features considering all the described architectures (Section~\ref{ssec:underlay-sp-analysis}). Lastly, we compare the underlay architectures (Section~\ref{ssec:underlay-comparison}).
%
%
%
\subsection{Description}
\label{ssec:underlay-description}
\subsubsection{\textbf{IICN~\cite{iicn}}}
\begin{figure}[!ht]
\centering
\includegraphics[width=0.9\linewidth]{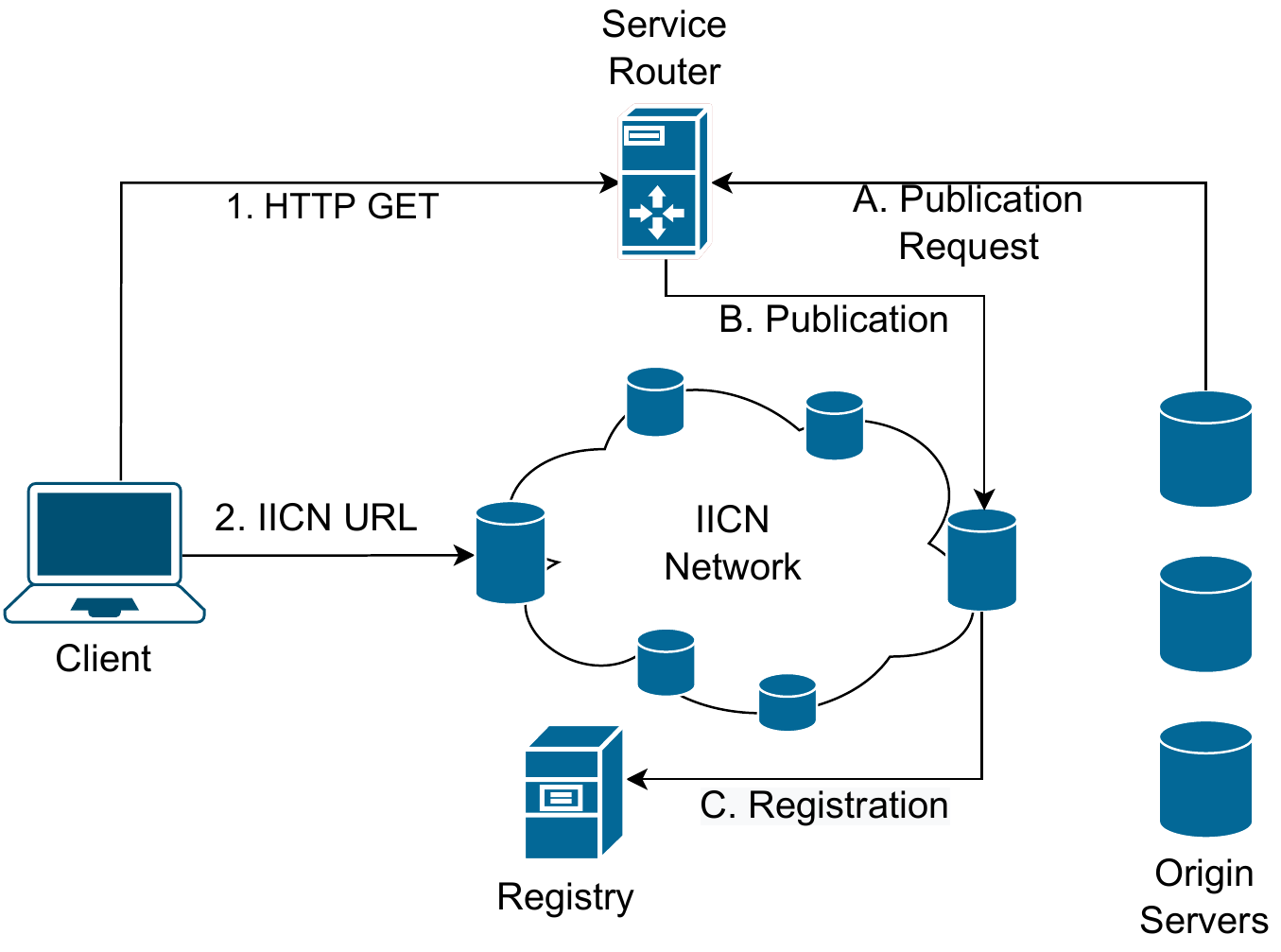}
\caption{The IICN~\cite{iicn} architecture. IICN network encompasses the ICN nodes that are enabled to retrieve information from IP origin servers via the service router and registry nodes.}
\label{fig:iicn_com}
\end{figure}

combines \gls{CDN} and \gls{ICN} concepts following the pull approach model as represented in Fig.~\ref{fig:iicn_com}. 
A client issues an interest request following the current IP paradigm, and a producer responds to this request.
The content can be either retrieved from the ICN network or the origin servers. 
Similar to \gls{CDN} networks, in IICN, there is the need for a service router that maps the requests to the location of the surrogate router -- i.e., ICN nodes -- from where the content can be fetched. 
Upon receiving an \gls{HTTP} GET request (Step 1), the service router converts this request into a set of identifiers -- i.e., routing identifiers and other sub-identifiers (Step 2). The modified request is forwarded to the nearest IICN node. 
The IICN nodes consult their \gls{FIB} tables to forward hop-by-hop the Interest request packet based on the routing information on the Interest packet. 
Furthermore, IICN introduces the registry node that stores and manages the mappings between information identifiers and their locations. 
Lastly, whenever the origin servers express intention to publish new content, it issues a request to the service router (Step A). After that, the service router sends the publication results -- i.e., mapping of \gls{URL} and identifiers -- towards the ICN node in the IICN network (Step B). The latter sends such registration results to the registry (Step C). 
IICN covers the ICN islands in IP ocean deployment scenario.

\subsubsection{\textbf{DOCTOR~\cite{doctor}}}
\begin{figure}[!ht]
\centering
\includegraphics[width=\linewidth]{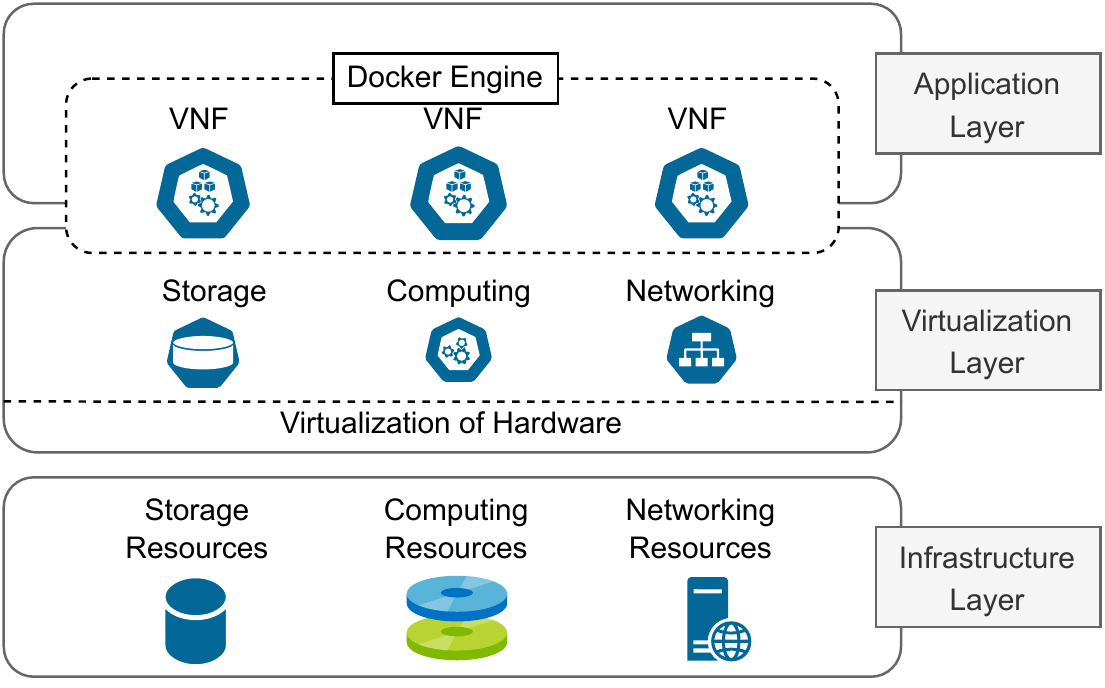}
\caption{The architecture of a DOCTOR~\cite{doctor} virtualized node. Here, three main layers are present---i.e., infrastructure, virtualization and application layers.}
\label{fig:doctor_node}
\end{figure}

proposes an NFV approach for NDN networks by considering Docker~\cite{boettiger2015introduction} as the central technology for implementing the \gls{NFV} infrastructure. 
The presence of the centralized model, similar to \gls{SDN}, allows the establishment of software control over the network. 
To accommodate \gls{NDN} protocol stack and deal with both types of traffic -- i.e., \gls{ICN} and \gls{IP} -- the \gls{NFV} framework is modified.
The virtual network proposed by DOCTOR is deployed based on the OpenSwitch to ensure the end-to-end connectivity between the virtualized network services.
The \gls{NDN} protocol is dockerized, creating a \gls{VNF}, and it is used both over \gls{IP} and Ethernet. Furthermore, to connect the \gls{ICN} islands through \gls{IP} infrastructure, DOCTOR introduces the ingress gateway (iGW) and egress gateway (eGW). The former translates the \gls{HTTP} requests to \gls{NDN} Interest requests, while the latter converts \gls{NDN} interests to \gls{HTTP} requests and \gls{HTTP} replies to \gls{NDN} data messages. Both gateways are virtualized, creating a \gls{VNF} for each of them, allowing ease of implementation anywhere in the network. Fig.~\ref{fig:doctor_node} represents the internal architecture of a DOCTOR virtualized node.
Given the design of DOCTOR, we consider it to cover all the deployment scenarios.
Lastly, DOCTOR architecture was proposed under the \textit{DeplOyment and seCurisaTion of new functiOnalities in virtualized networking enviRonnements} project  French Nation Research Agency that started in 2014 and ended in 2018.

\subsubsection{\textbf{POINT~\cite{point}}}
\begin{figure}[!ht]
\centering
\includegraphics[width=\linewidth]{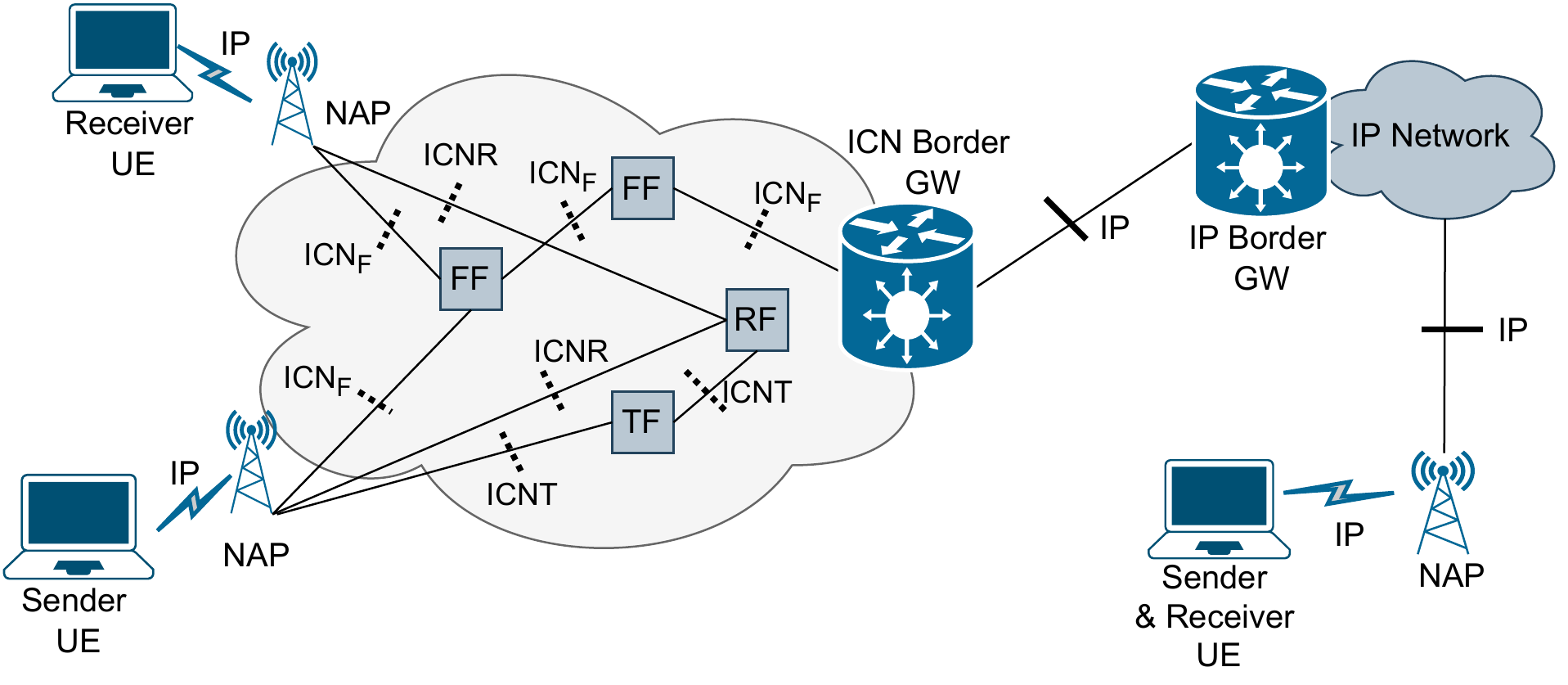}
\caption{The POINT~\cite{point} architecture. In its core POINT includes the Topology Function (TF), Forwarding Function (FF) and the Rendezvous Function (RF). Additionally, the border gateways interconnect IP and ICN islands. The doted lines represent the communication interfaces.}
\label{fig:point_com}
\end{figure}

is  an evolution of PURSUIT~\cite{pursuit} since it leverages functions such as \gls{RF}, \gls{TF} and \gls{FF}. 
POINT exploits the \gls{SDN} paradigm that allows using standard Ethernet switches. 
To provide support for \gls{IP} based applications to run in an \gls{ICN} setup, POINT introduces an abstraction layer that allows existing applications to use \gls{ICN} parts of the network without changing their application interface. 
Here, it follows a gateway-based approach, where the first hop of communication from the end user towards the network is established through \gls{IP} interface using \gls{HTTP} or \gls{CoAP}.
\gls{NAP} is the entity that maps the IP-based protocol abstraction to \gls{ICN} semantics. Furthermore, POINT adopts the \gls{ICN} border gateway, a new entity that allows the communication between \gls{IP} and \gls{ICN} networks. 
Fig.~\ref{fig:point_com} represents the communication entities and interfaces.
Lastly, POINT introduces SDN and ICN-over-SDN layers on top of the link layer to exploit the SDN functionality. 
Therefore, POINT covers the border island deployment scenario.
POINT is evaluated using a Blackadder ICN platform developed within PURSUIT~\cite{pursuit} and OpenFlow protocol~\cite{hu2014survey}. 
The former enables ICN functions, including RF, TF, and FF. The Blackadder platform uses the OpenFlow protocol to allow ICN routing, and the flow rules are indicated by the controller---i.e., the topology manager.
Lastly, POINT architecture was proposed under the \textit{H2020 project iP Over IcN- the betTer IP (POINT)} that started in January 2015 and ended in December 2017. All the project source code\footnote{ \url{https://github.com/point-h2020}} is available and can be used to carry more research on the underlay coexistence approach.

\subsubsection{\textbf{RIFE~\cite{rife}}}
-- i.e., architectuRe for an Internet For Everybody -- combines together \gls{DTN}, \gls{IP} and \gls{ICN}. 
\gls{DTN} is related to the constrained communication environments due to the added delay or disruption tolerance. The application of the \gls{ICN} benefits in a \gls{DTN} network allows reducing further costs on accessing the content by caching it on the edge of the network and transmitting it only upon request.
The design of RIFE exploits concepts introduced by PURSUIT since its graph-based object model integrates well with \gls{DTN}. \gls{RF}, \gls{TF} and \gls{FF} are present in RIFE as well, and similarly to \gls{SDN}, the forwarding plane is decoupled from the data plane. 
Same as POINT, RIFE exploits the gateway approach by introducing \gls{NAP} that allows end-users to communicate to the network, and RIFE Border GW, which provides access to the public Internet. Due to the similarity with POINT architecture, the communication flow follows the same logic. 
Here, a node encompasses a dissemination strategy layer and supports different network interfaces, including physical Network Interface Cards (NIC) and logical network interfaces. Dissemination strategies and routing tables govern the communication flow between and inside nodes.
Lastly, RIFE adheres to the border island deployment scenario.
RIFE architecture was design under the \textit{Horizon2020 architectuRe for an Internet For Everybody} project, which started in February 2015 and ended in January 2018.

\subsubsection{\textbf{CableLabs~\cite{cacheLabs}}} 
-- i.e., Cable Television Laboratories -- combines \gls{CDN} and \gls{ICN} in an underlying fashion and aims to take advantage of the benefits of ICN on CDN networks. 
Here, CableLabs introduces the HTTP-to-ICN and ICN-to-HTTP proxies that allow the implementation of \gls{ICN} and \gls{IP} islands in the \gls{IP} and \gls{ICN} oceans. 
Furthermore, it places caches in an in-path fashion -- i.e., on the path between consumers and origin servers -- overloading this way the origin servers.
Furthermore, CableLabs introduce proxies that ensure the transition from \gls{HTTP} based \gls{CDN} to \gls{ICN} based \gls{CDN}.
A combination of a \gls{HTTP} server and an \gls{ICN} consumer is used for the HTTP-to-ICN proxy. The server receives HTTP GET from a client and transforms it to an \gls{ICN} interest. While the ICN-to-HTTP proxy, it combines a CCN publisher and an HTTP client.
CableLabs upgrades a set of edge caches to support ICN forwarding and caching. Therefore, it places HTTP-to-ICN proxies in the north side connection of edge caches with other aggregation routers and ICN-to-HTTP proxies in the south side connection of edge caches with clients. 
%
%
The aggregation router maintains FIB with entries containing name prefixes that the caching infrastructure can serve. Each entry points to the cache's face if the request comes from a cache and to the consumer's face if the request came directly from a consumer. 
Lastly, CableLabs covers all the deployment scenarios except the border island scenario.
CableLabs architecture is designed by CableLabs\footnote{ \url{https://www.cablelabs.com/}}, which is an Innovation and R\&D lab focused on building and orchestrating emergent technology.
\subsubsection{\textbf{COIN~\cite{coin}}} 
\begin{figure}[!ht]
\centering
\includegraphics[width=0.9\linewidth]{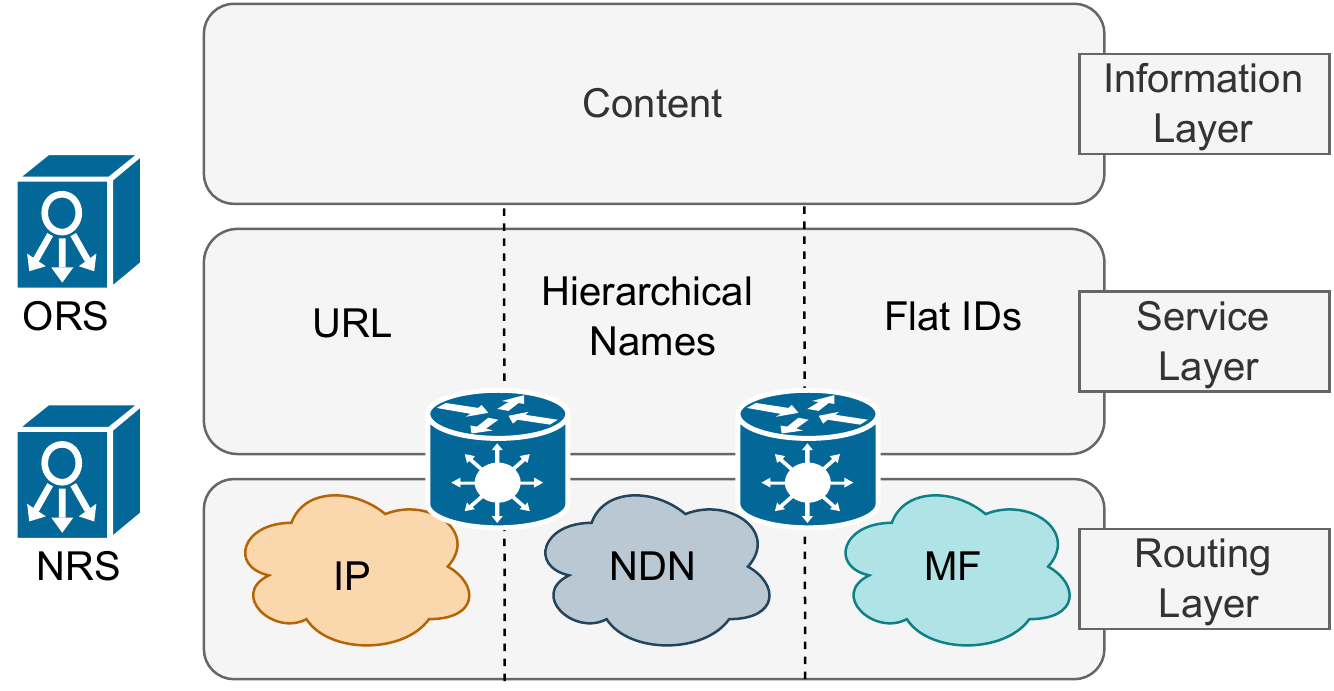}
\caption{The COIN~\cite{coin} architecture. COIN designs three layers -- i.e., routing, service and information layer -- that enable the interoperability of IP, NDN adn MF. Additionally the resolution service nodes assist for content retrieval from these domains.}
\label{fig:coin_architecture}
\end{figure}

aims to offer interoperability for current and future -- i.e., NDN and MobilityFirst (MF)-- Internet architectures.
%
%
The coexistence among different domains in COIN is achieved by employing translation of content flowing between domains. The translation is performed on the content-level layers---i.e., network layer for \gls{ICN} and application layer for \gls{IP} networks. 
For this purpose, it uses gateways that track state information of requests and responses to achieve this aim. 
COIN architecture comprises three layers: the information layer that captures content items in different domains, the service layer that indicates the name format for each content item on the information layer, and the routing layer that is in charge of transmitting the packets correctly, as shown in Fig.~\ref{fig:coin_architecture}. 
Furthermore, COIN introduces the \gls{NRS} nodes for static and dynamic content retrieval in multiple heterogeneous domains. 
Naming schemes remain unchanged for all domains -- e.g., \gls{NDN} keeps its hierarchical naming -- and the consumer issues requests using the "ContentDomain + ContentName" expression. This information can be retrieved from the \gls{NRS} nodes. 
If a consumer from one domain -- e.g., \gls{IP} domain -- requests content from another domain -- e.g., \gls{NDN} domain -- the gateway behaves like the end of \gls{TCP} communication for the requests that come from the \gls{IP} domain and an \gls{NDN} client for the \gls{NDN} domain. The gateway stores the state information needed to route back the response---i.e., the consumer's IP address and source port. 
For evaluation, COIN uses the CCNx library to implement the NDN components, MobilityFirst project~\cite{MF} and basic Linux implementation of the IP forwarding.
COIN covers all the deployment scenarios.

\subsection{SP Analysis}
\label{ssec:underlay-sp-analysis}

\subsubsection{\textbf{Trust}}  
Almost all the analysed underlay architectures contain both IP and ICN or NDN islands in their deployment scenarios.
Generally, in \gls{IP} islands, trust is based on securing communication channels, while in \gls{ICN} islands, trust is based on the content itself. Therefore, the considerations made for the trust feature of the overlay architectures also suit the underlay deployment.
However, the majority of the analysed architectures -- i.e., \textbf{IICN}, \textbf{DOCTOR}, \textbf{CableLabs}, \textbf{POINT} and \textbf{RIFE} -- do not clearly describe the trust mechanisms in such islands or in the communication between them. 
Additionally, the underlay architectures often use the gateways to realise the communication among different domains. 
In this context, \textbf{COIN} requires the gateway to decrypt the information coming from an \gls{IP} domain to get the content name and other features that are necessary for information retrieval. Afterward, the gateway encrypts only the payload for content from content-based domains and leaves all headers in clear. This way, the gateway only modifies the headers without decrypting the payload.
Even in this case, the needed infrastructure that enables key sharing in different domains is not elaborated during the architecture's design.

\subsubsection{\textbf{Data origin authentication}}
Considering that the origin servers sign each content in ICN islands of \textbf{IICN}, data origin authentication can be achieved. 
Furthermore, tunnel mode must be used during the communication between different ICN and IP islands.
Instead, for \textbf{DOCTOR}, \textbf{POINT}, \textbf{RIFE} and \textbf{CableLabs} that include rather the border island or all scenarios deployment, an additional effort is needed to ensure origin authentication for the data flowing in different islands and oceans.
Lastly, \textbf{COIN} resolves the interoperability problem by exploiting transitive trust~\cite{transitive-trust} and the gateways. Here, the producer's signature in one domain is verified by the gateway situated on the producer's side and only afterward signs the data with its private key. Therefore, the consumer proves only the last hop gateway.

\subsubsection{\textbf{Peer entity authentication}}  
All the analysed underlay architectures include both IP and ICN islands in their deployment scenarios. 
Therefore, \textbf{IICN}, \textbf{DOCTOR}, \textbf{CableLabs} and \textbf{COIN} lack peer entity authentication in the ICN islands. Instead, existing mechanisms that ensure peer entity authentication can be used for the IP islands.
Lastly, since \textbf{POINT} \textbf{RIFE}, \textbf{COIN} and \textbf{DOCTOR} include the border island deployment scenario, they must also consider the presence of entities present in different oceans rather than considering only the islands. 

\subsubsection{\textbf{Data integrity}}  
The same rationale of the data origin authentication analysis follows also for data integrity of \textbf{IICN}, \textbf{DOCTOR}, \textbf{POINT}, \textbf{RIFE} and \textbf{CableLabs}. 
Instead, \textbf{COIN} elaborates that cryptographic hash functions -- e.g., MD5, SHA-1, SHA-256 -- can be used and applied to the data and announced to consumers by the producers. 
Here, the gateways are a crucial component in establishing this feature between different domains since they resign the content from another domain with the private key of the following domain. Therefore, the consumer verifies and trusts the last hop gateway.

\subsubsection{\textbf{Authorization and access control}} 
In IP islands for all the architectures, given the presence of \gls{IP} addresses during the communication, authorization and access control can be achieved using \gls{ACL} constructed on hosts' \gls{IP} addresses. Nevertheless, in such islands, countermeasures against attacks -- e.g., IP spoofing -- should be used not to disrupt the rules of ACL.
Instead, a mechanism to deny or grant access to nodes is not provided for ICN islands. 
Additionally, the architectures that introduce gateway nodes -- i.e., \textbf{DOCTOR}, \textbf{POINT}, \textbf{RIFE} and \textbf{COIN} -- might exploit them to establish access permissions based both on content names and \gls{IP} addresses.

\subsubsection{\textbf{Accountability}}
For all the underlay architectures, in \gls{ICN} islands, only origin content producers can be traced, assuming they sign each generated content.
Instead, in IP islands, the presence of \gls{IP} addresses permits the possibility of tracing the actions in the network. 
However, existing attacks such as \gls{MAC} or \gls{IP} spoofing might disturb the achievement of such \gls{SP} feature.
\textbf{DOCTOR}, \textbf{POINT}, \textbf{RIFE} and \textbf{COIN} might apply accountability procedures on the gateways.
Here, for communication flowing from \gls{ICN} to \gls{IP} islands, the gateways can trace both \gls{IP} addresses and requested names.

\subsubsection{\textbf{Data confidentiality}} 

We consider data confidentiality not provided for almost all the analysed architectures, except for \textbf{COIN}. 
In particular, \textbf{IICN}, \textbf{DOCTOR}, \textbf{POINT}, \textbf{RIFE} and \textbf{CableLabs} that adhere to border island or all scenarios deployment, must design accurate encryption schemes and procedures that deal with domain heterogeneity.
Instead, \textbf{COIN} designs and motivates the encryption scheme. On the one hand, if the communication is between two \gls{NDN} islands -- i.e., both consumer and producer have content-oriented security -- they sign the requests and data using their cryptographic keys. In this case, the gateways are "blind" -- i.e., they do not encrypt or decrypt the packets -- but only translate from one domain namespace to the other. On the other hand, if the communication is established between an \gls{IP} host and \gls{NDN} host, the gateways need to re-encrypt the data received in one domain to provide data confidentiality while delivering the content to the other domain.

\subsubsection{\textbf{Availability}}
Both \textbf{IICN} and \textbf{CableLabs} architectures leverage CDN features during their design.
Triukose et al.~\cite{triukose2009content} have shown that CDN networks are vulnerable to DoS attacks. In \textbf{IICN}, all the routers are \gls{IP} legacy routers and inherit all IP availability issues. In addition, the registry node, which contains the mappings of information identifiers used during forwarding and information location, can be the target of attacks that aim to modify such mappings and deny access to the content.
Instead, all caches -- i.e., edge and mid caches -- in \textbf{CableLabs} are similar to ICN caches and can be the target of ICN cache attacks---i.e., cache pollution, content poisoning. Furthermore, by attacking the proxies -- i.e., HTTP-to-ICN and ICN-to-HTTP -- the translation of HTTP GET requests to ICN names can be disrupted. In addition, the presence of FIB tables on aggregation routers can be exploited by attackers that target such resources to deny the regular communication flow.
The gateways that several analysed architectures -- i.e., \textbf{DOCTOR}, \textbf{POINT}, \textbf{RIFE} and \textbf{COIN} -- use to enable the communication between different islands, present a single point of failure.
For example, in \textbf{POINT} and \textbf{RIFE}, the \gls{NAP} nodes are fundamental since they enable the translation from one protocol semantic to the other but also a new target for possible new availability attacks.
\textbf{DOCTOR} presents different \gls{VNF} to implement the \gls{NDN} and \gls{IP} stacks. \gls{DoS} attacks may be directed to virtual networks or VNFs' public interfaces to exhaust network resources and impact service availability~\cite{lal2017nfv}. Furthermore, a huge traffic volume from a compromised VNF can be generated and sent to other VNFs.

\subsubsection{\textbf{Anonymous communication}}  
The ICN islands in \textbf{IICN}, \textbf{DOCTOR}, \textbf{CableLabs} and \textbf{COIN} architectures, ensure consumers anonymity. Instead, all the IP islands lack communication anonymity.
Additionally, in \textbf{IICN}, the service routers might leak information related to the identity of IICN nodes located in the \gls{ICN} islands. 
The state tables present on the gateways in \textbf{DOCTOR}, \textbf{POINT}, \textbf{RIFE} and \textbf{COIN}, might be used to deanonymise the communication parts.

\subsubsection{\textbf{Traffic flow confidentiality}}
Both the \gls{IP} and \gls{ICN} islands inherit known related traffic analysis issues in the analyzed architectures.
These attacks can be mitigated by implementing the already proposed countermeasures. However, only a few countermeasures have been proposed for the \gls{ICN} islands to mitigate such attacks.
\textbf{IICN} is the only underlay architecture that tries to reduce traffic confidentiality issues. Here, the \gls{ICN} islands do not use the hierarchical naming schema as in native \gls{ICN} but map the URLs into routing identifiers.


\subsection{Comparison}
\label{ssec:underlay-comparison}

Table~\ref{tab:underlay-evaluation} depicts the evaluation scores for the analysed underlay architectures. The achievement of \gls{SP} features in a combination of heterogeneous domains is not a trivial task due to different security models and mechanisms.
As this table shows, \textbf{COIN} attains averagely the highest scores. Indeed, during its design, COIN focuses on two essential security mechanisms -- i.e., encryption and signatures -- ensuring trust, data confidentiality, origin authentication, and integrity.
Nevertheless, similarly to the other described architectures, COIN lacks peer entity authentication. Furthermore, it fails to mitigate availability issues and anonymous communication and traffic flow confidentiality mainly due to the presence of gateways.
After COIN, \textbf{IICN} scores the highest average. For features such as data origin authentication and integrity, accountability, and anonymous communication, it mostly takes advantage of the enhanced security features of ICN islands. 
Lastly, \textbf{POINT}, \textbf{RIFE}, \textbf{DOCTOR} and \textbf{CableLabs} fail to ensure different features mostly due to their border island or all scenario deployment. Such deployment scenarios are challenging, and only appropriately designed mechanisms can ensure SP features.
Instead, for what concerns the feature coverage for the analysed underlay architectures, \textbf{trust} and \textbf{accountability} are amongst the mostly achieved features, while \textbf{peer entity authentication}, \textbf{data confidentiality}, \textbf{availability}, \textbf{anonymous communication} and \textbf{traffic flow confidentiality} the less covered features. 
Indeed, we have considered the gateways in almost all the underlay architectures crucial for the accountability feature. These nodes can be used for accountability procedures in communication among different domains.
\begin{table}[!htb] 
\centering
\caption{Evaluation of security and privacy features of underlay IP-ICN coexistence architectures.} 
\label{tab:underlay-evaluation}
\resizebox{0.7\columnwidth}{!}{
    \begin{tabular}{c | c c c c c c}
      \toprule
      \makecell{\textbf{SP Feature}} & 
      \makecell{\textbf{\rotatebox{90}{\scriptsize IICN~\cite{iicn}}}} & 
      \makecell{\textbf{\rotatebox{90}{\scriptsize DOCTOR~\cite{doctor}}}} & 
      \makecell{\textbf{\rotatebox{90}{\scriptsize POINT~\cite{point}}}} &
      \makecell{\textbf{\rotatebox{90}{\scriptsize RIFE~\cite{rife}}}} &
      \makecell{\textbf{\rotatebox{90}{\scriptsize CableLabs~\cite{cacheLabs}}}} &
      \makecell{\textbf{\rotatebox{90}{\scriptsize COIN~\cite{coin}}}} \\
      \toprule 
       \makecell{ Trust} & \tiny \pie{180} & \tiny \pie{180} & \tiny \pie{180} & \tiny \pie{180} & \tiny \pie{180} & \tiny \pie{360} \\ 
        \midrule
        \makecell{\thead{Data origin \\ authentication}}  & \tiny \pie{180} & \tiny \pie{0} & \tiny \pie{0} & \tiny \pie{0} & \tiny \pie{0} & \tiny \pie{360} \\ 
        \midrule
        \makecell{Peer entity \\ authentication} & \tiny \pie{180} & \tiny \pie{0} & \tiny \pie{0} & \tiny \pie{0} & \tiny \pie{180} & \tiny \pie{0} \\ 
        \midrule
        \makecell{Data integrity}  & \tiny \pie{180} & \tiny \pie{0} & \tiny \pie{0} & \tiny \pie{0} & \tiny \pie{0} & \tiny \pie{360} \\ 
        \midrule
        \makecell{Authorization \\ \& access control} & \tiny \pie{0} & \tiny \pie{180} & \tiny \pie{180} & \tiny \pie{180} & \tiny \pie{0} & \tiny \pie{180} \\ 
        \midrule
        \makecell{Accountability} & \tiny \pie{180} & \tiny \pie{180} & \tiny \pie{180} & \tiny \pie{180} & \tiny \pie{1800} & \tiny \pie{180} \\ 
        \midrule
        \makecell{Data \\ confidentiality} & \tiny \pie{0} & \tiny \pie{0} & \tiny \pie{0} & \tiny \pie{0} & \tiny \pie{0} & \tiny \pie{360} \\ 
        \midrule
        \makecell{Availability} & \tiny \pie{0} & \tiny \pie{0} & \tiny \pie{0} & \tiny \pie{0} & \tiny \pie{0} & \tiny \pie{0} \\ 
        \midrule
        \makecell{Anonymous \\ communication} & \tiny \pie{180} & \tiny \pie{0} & \tiny \pie{0} & \tiny \pie{0} & \tiny \pie{180} & \tiny \pie{0} \\ 
        \midrule
        \makecell{Traffic flow \\ confidentiality} & \tiny \pie{180} & \tiny \pie{0} & \tiny \pie{0} & \tiny \pie{0} & \tiny \pie{0} & \tiny \pie{0} \\ 
        \bottomrule
    \end{tabular}}
\\[1\baselineskip]
\tiny \pie{360} \scriptsize Fulfilled~~~~~~ \tiny \pie{180} \scriptsize{Partially fulfilled}~~~~~~\tiny \pie{0} \scriptsize{Not fulfilled}
\end{table}

\section{Hybrid Architectures}  
\label{sec:hybrid}
%
This section provides the \gls{SP} analysis for the architectures adhering to the hybrid deployment approach. Generally, such architectures adopt dual-stack switching nodes which can handle the semantics of both \gls{IP} and \gls{ICN} packets. 
We describe the hybrid architectures (Section~\ref{ssec:hybrid-description}), and then we provide the SP analysis for the depicted architectures (Section~\ref{ssec:hybrid-sp-analysis}). 
We conclude this section by comparing the described architectures (Section~\ref{ssec:hybrid-comparison}).
\subsection{Description}
\label{ssec:hybrid-description}

\subsubsection{\textbf{NDN-LAN~\cite{ndnLAN}}}
\begin{figure*}[!ht]
\centering
\includegraphics[width=0.75\linewidth]{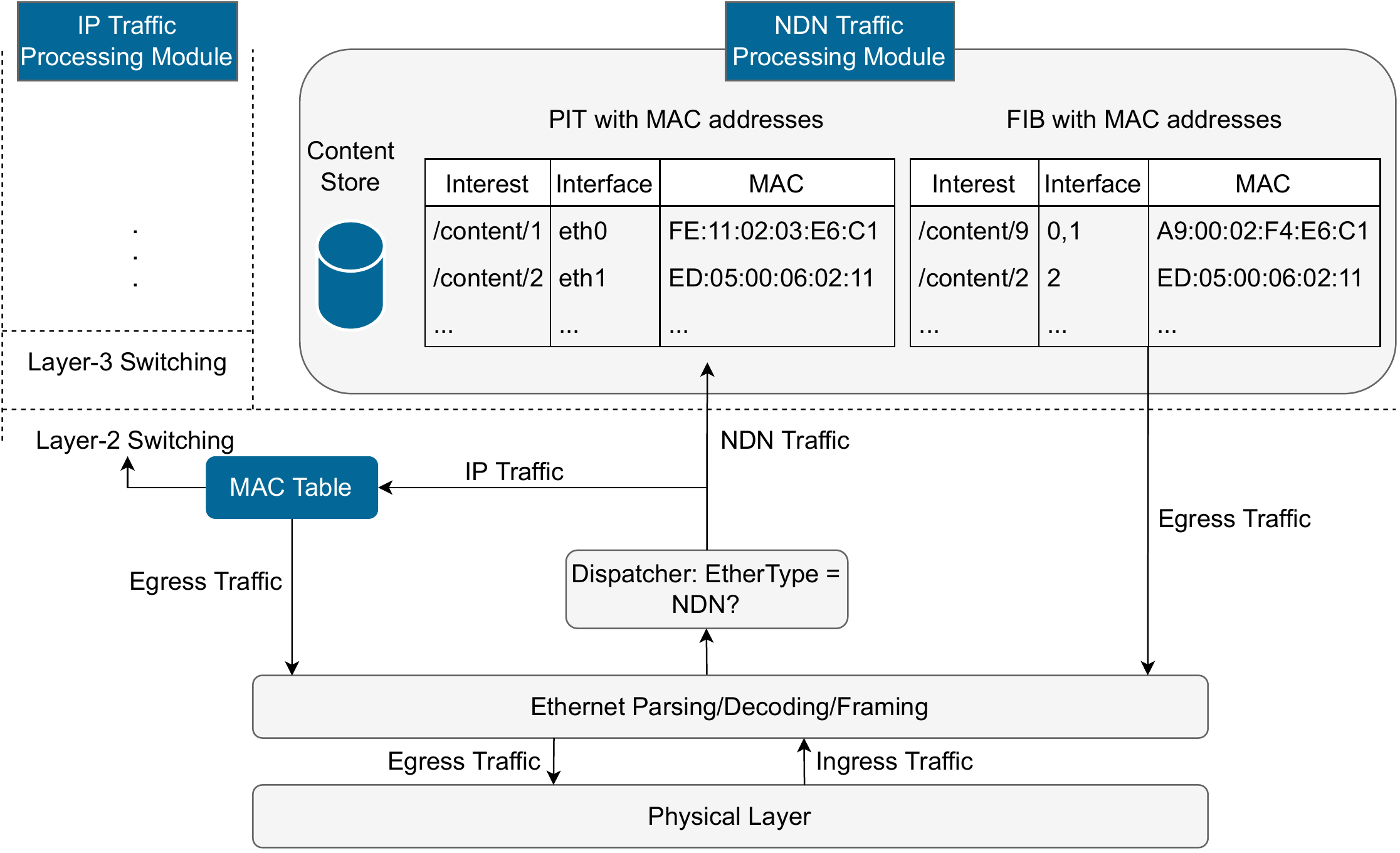}
\caption{NDNLAN~\cite{ndnLAN} Dual-Stack switch internal structure. The dispatcher captures the traffic type -- i.e., IP or NDN -- and dispatches it accordingly. While the IP traffic follows standard processing, the NDN traffic is processed in a dedicated module composed of dedicated PIT and FIB.}
\label{fig:dualstack}
\end{figure*}

is a hybrid architecture that can process both IP and ICN traffic. 
It implements \gls{NDN} protocol over the Ethernet by mapping the contents to \gls{MAC} addresses. 
Depending on the network configuration, NDN-LAN can use a Dual-Stack switch -- i.e., D-switch -- Ethernet switch -- i.e., E-switch -- and \gls{NDN} switch---i.e., N-switch. 
The former is a hybrid switch that forwards the \gls{IP} packets based on \gls{IP} addresses and the \gls{NDN} packets based on the name prefixes.
Conversely, E-switch and N-switch only support host-based and name-based forwarding, respectively. 
The implementation scenarios for NDN-LAN are: (a) NDN or IP enabled hosts and only E-switches, (b) NDN or IP hosts and only D-switches, and (c) hybrid network with both E-switches and D-switches. 
In the first case, the hosts must maintain a forwarding table of mappings between the name prefix and a destination \gls{MAC} address. 
Instead, in the second case, the D-switches maintain associations of name prefix, incoming interface, and \gls{MAC} address.
This trio is stored in a FIB entry. Furthermore, the D-Switch checks the EtherType in the Ethernet header for identifying the \gls{NDN} traffic.
The D-switch forwards the IP traffic based on the destination \gls{MAC} address, while the NDN traffic is delivered based on the names carried in the NDN packet. It builds its FIB by self-learning mechanism---i.e., in case of a FIB miss, it sends the interest request to all interfaces except the one from which the request is received. Fig.~\ref{fig:dualstack} depicts the internal structure of the D-switch.
Lastly, the hybrid case can either forward the \gls{NDN} packets based on name prefixes by D-switches or MAC address by E-switch. However, the network should be carefully designed to avoid name-based and host-based forwarding conflicts. 
NDN-LAN covers all the deployment scenarios expect the border island scenario.

\subsubsection{\textbf{hICN~\cite{hICN}}}
\begin{figure*}[!ht]
\centering
\includegraphics[width=0.7\linewidth]{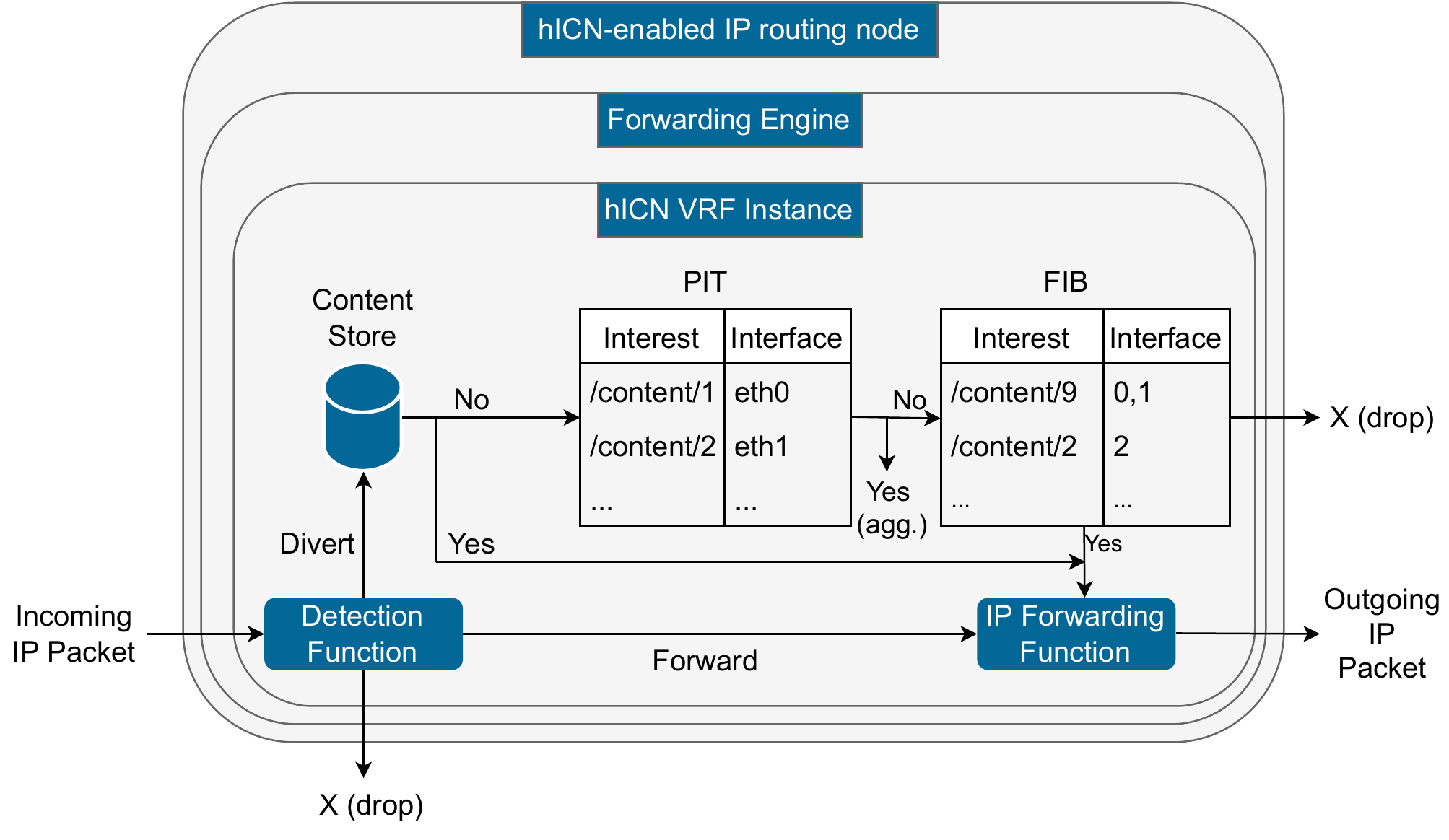}
\caption{The internal structure of an hICN~\cite{hICN} routing node. Similarly to D-switch, the CS, PIT and FIB instances are arranged inside the node for processing the NDN traffic.}
\label{fig:hICNnode}
\end{figure*}

is a proposal of Cisco that aims to integrate \gls{ICN} semantics inside the \gls{IP} protocol while making use of the benefits that \gls{ICN} proposes.
Such integration is not achieved either using tunneling or encapsulation. Instead, it is designed to make both types of traffic cohabit in the same infrastructure. 
The principal components in hICN are a) hICN routers, b) IP routers, and c) hICN-enabled IP routers. The latter category includes routers capable of processing and forwarding regular IP and ICN-enabled IP packets.
In hICN, data is referred to by name, containing name prefix and name suffix. The former is similar to the \gls{IP} addresses since routers use it during forwarding. Instead, the latter contains mainly segmentation information. For data packets, the name prefix is accommodated to the source \gls{IP} address, while for the interest packets, it is placed to the destination \gls{IP} address. Conversely, the name suffix is placed in the \gls{TCP} sequence number field. The \gls{FIB} tables used during forwarding contain IP addresses and name prefixes. The \gls{ICN} forwarders are equipped with the \gls{PIT} and \gls{CS}. Instead, the normal \gls{IP} routers encompass both these components in the so-called packet cache, which can store both packets. Fig.~\ref{fig:hICNnode} presents the internal structure of an hICN node.
The authors have evaluated both the feasibility assessment and the performance of hICN. Nevertheless, thy do not provide detailed descriptions regarding the hICN protocol implementation.
Lastly, hICN covers all the deployment scenarios.
The source code\footnote{\url{https://github.com/FDio/hicn}} for the implementation of Cisco's hICN is available and maintained by an active group of contributors. Therefore, it can be used to experiment on the hybrid coexistence approach.

\subsubsection{\textbf{OFELIA~\cite{ofelia}}}  
is an extension of CONET~\cite{conet}  designed for OpenFlow networks under the OFELIA project.
Similarly to CONET, the participating nodes in the communication are BN, IN, SN, \gls{NRS} and end nodes. 
In OFELIA, the content names are carried in the \gls{IP} option and are mapped to a 4-bytes long tag which can be inspected in the OpenFlow network.
Such mapping is managed by the \gls{NRS} which ensures that it is identical for all the BN. Only the first content request is subject to mapping request to the \gls{NRS} node.
Upon receiving a packet, the BN of an OpenFlow CSS checks the type of packet---i.e., IP packet or CONET packet. 
The IP packets are managed by the network using standard OpenFlow mechanisms. Conversely, for CONET packets, a different packet processing is followed. 
The interest packets are forwarded to the destination passing through different hops of BN.
Here, the controller assigns a flow identifier for the interest that the BN uses to forward the interest to the SN. 
Instead, in the downstream path, the BN encapsulates the named data in an IP packet and sets the destination IP address to the address of the next BN. The content can be cached on IN situated in the path between BN. 
Lastly, OFELIA adheres to the border island deployment scenario.

\subsubsection{\textbf{CLIP~\cite{clip}}}
is a hybrid architecture that associates a globally unique publisher label to a unique content item label. 
The labels assigned to the publishers in CLIP can be global, local, or anonymous. 
The former is 64 bytes long to be compliant with the smallest allocation of IPv6 address space and can be globally recognizable. 
Conversely, the local publisher label is usually assigned by the network provider and manages the locally produced and consumed data. Instead, the anonymous publisher labels are used in case of one-time data transfer or hiding the proper publisher. 
CLIP creates a large ICN subnet containing all the \gls{ICN} traffic, with individual subnets for each publisher. 
The content names are integrated into the IPv6 address space, particularly in the destinations option header field in the IPv6 datagram. 
Lastly, CLIP covers all the deployment scenarios expect the border island scenario.
\subsection{SP Analysis}
\label{ssec:hybrid-sp-analysis}
\subsubsection{\textbf{Trust}}  
Generally, trust in hybrid architectures must include both IP and ICN trust models. Indeed, \textbf{NDN-LAN} and \textbf{CLIP} include all but border island deployment scenarios, while \textbf{hICN} includes all deployment scenarios and \textbf{OFELIA} border island deployment scenario.
In particular, \textbf{NDN-LAN} introduces a way to build hybrid Local Area Networks (LAN) with the presence of a Dual-switch that processes both \gls{NDN} and \gls{IP} traffic. Furthermore, the D-switch is responsible for maintaining the mapping between name prefixes and destination MAC addresses to ensure the forwarding of NDN traffic. Even though NDN-LAN provides means to achieve information-centric benefits -- i.e., routing by name, consumer anonymisation, and caching -- the producer is still reached based on its \gls{MAC} address. Therefore, NDN-LAN must design trust schemes that consider both channel-based and content-based trust.
\textbf{CLIP} discusses and motivates its trust scheme based on a one-way relationship between hosts. All the datagram elements -- except for the traffic class, flow label, and hop limit -- are signed here. Moreover, the authors affirmed that a self-certifying naming scheme for implementing security features is not recommended due to several vulnerabilities---e.g., key compromises or \gls{DoS}.
Similar to \textbf{CLIP}, \textbf{h-ICN} explicitly describes trust scheme. Here, trust is based on the network service, similar to DHCPv6, which leases the hosts' prefixes. In particular, whenever a producer requests a network prefix, it must send its public key and identity to the network service. Upon receiving such a request, the network service will create and sign with its private key the hICN data, including the producer's public key, identity, producer's prefix, and the lease time. The name of such data is used as a key locator for the producer's public key.
Lastly, \textbf{OFELIA} encompasses both types of traffic -- i.e., IP and CONET -- and decouples trust on both host and content. For the regular IP traffic, standard management trust mechanisms are followed. Instead, for CONET traffic, the content itself is secured. Additionally, OFELIA engages the SDN concept by using multiple OpenFlow controllers. Such controllers can play an essential role in establishing trust in the network.

\subsubsection{\textbf{Data origin authentication}}
All the hybrid architectures ensure data origin authentication in the ICN islands. Conversely, IPSec must be used in tunnel mode to provide such a feature in IP islands.
In particular, since NDN traffic is encapsulated in the IP packets by modifying the Layer-2 header of the packet, data origin authentication in \textbf{NDN-LAN} is ensured in case not only the content is signed but also the \gls{IPSec} is used. 
Additionally, \textbf{CLIP} relies on the \gls{AH} and \gls{ESP} protocols of \gls{IPSec} in order to provide data origin authentication.
Instead, both \textbf{OFELIA} and \textbf{hICN} create \gls{IP} and \gls{ICN} packet flows.
For regular \gls{IP} packets, assuming that \gls{IPSec} protocol is used, these architectures guarantee data origin authentication due to the presence of \gls{AH} and \gls{ESP} protocols. Instead, consumers can check such a feature for the ICN packets by validating the producer's signature. 
Nevertheless, in the communication between different oceans of \textbf{hICN} and \textbf{OFELIA}, they must also deal with varying models of security --i.e., host-based in IP and content-based in ICN -- to ensure data origin authentication feature. 
%
%
%

\subsubsection{\textbf{Peer entity authentication}} 
All the ICN islands of the analysed hybrid architectures lack peer entity authentication.
Instead, for the \gls{IP} islands, existing peer entity authentication mechanisms might be used. For example, in \textbf{CLIP}, the authors claim that the associations between real-world identities and the publisher's label are established by creating a \gls{SA} and using protocols such as \gls{ISAKMP} and IKE. 

\subsubsection{\textbf{Data integrity}} 
Following the same rationale for data origin authentication, data integrity is ensured for all the ICN islands of the analysed architectures.
Indeed, \textbf{hICN} and \textbf{OFELIA} still have to face the open issue of different oceans involved during the communication of different islands. 
Instead, \textbf{NDN-LAN} and \textbf{CLIP} rely on the AH and ESP protocols of IPSec to ensure data integrity.
%
%
%
%

\subsubsection{\textbf{Authorization and access control}} 
For \textbf{NDN-LAN}, \textbf{hICN} and \textbf{CLIP}, authorization and access control can be achieved through the use of \gls{ACL} that constructed based on IP addresses. Alternatively, the name prefixes can also be used to create control over the namespaces.
Additionally, since \textbf{OFELIA} adheres to the border island implementation, it can exploit BN nodes and apply ACL built on both content and hosts, similar to the egress routing on IP networks. Here, the rules for updating and managing the ACL can be delegated to the OpenFlow controller.
Nevertheless, a trade-off must be found since the BN nodes and OpenFlow controller represent a single failure point for OFELIA.

\subsubsection{\textbf{Accountability}}  
In all ICN islands, the presence of signatures in data packets ensures accountability for producers. Furthermore, the presence of mappings between content names and \gls{MAC} addresses serving this content in D-switches of \textbf{NDN-LAN} further ensures accountability for producers.
On the other hand, almost all the architectures modify the IP packets to accommodate the ICN semantics. Therefore, the communication is still established using IP addresses, allowing the accountability procedures.
Nevertheless, such procedures can also be the subject of spoofing attacks. Egress filtering on different access nodes can be applied to ensure that the outgoing traffic originates from the addressed part of the network.

\subsubsection{\textbf{Data confidentiality}} 
None of the analysed hybrid architectures provide mechanisms and techniques to ensure data confidentiality.
Additionally, the authors of \textbf{hICN} claim that they delegate the data confidentiality feature to the upper layer -- i.e., secure transport -- and more research is needed, leaving space for future work. 

\subsubsection{\textbf{Availability}}
\textbf{NDN-LAN} and \textbf{CLIP} inherits existing \gls{IP} and \gls{ICN} availability vulnerabilities in the IP and ICN islands. Additionally, NDN-LAN introduces D-switches equipped with caches. 
Here, cache-related attacks -- i.e., content and cache pollution -- are inherited.
D-switches can also be the target of \gls{DoS} and \gls{DDoS} attacks that exhaust their resources.
Moreover, \textbf{hICN} enables routers to contain both IP addresses and name prefixes in their forwarding tables. Therefore, flooding and \gls{DoS} or \gls{DDoS} attacks might target such components. Furthermore, the presence of caches on such routers makes \textbf{hICN} inherit all the cache-related vulnerabilities of \gls{ICN}.
Similarly, \textbf{OFELIA} proposes hybrid FIB tables -- i.e., containing both name prefixes and IP addresses -- and OpenFlow controller. The latter is considered a single point of failure that might introduce control layer availability issues. Furthermore, for the IP CSS islands, OFELIA inherits already known availability issues of IP networks.

\subsubsection{\textbf{Anonymous communication}} 
\textbf{NDN-LAN} and \textbf{CLIP} inherit the lack of anonymous communication from \gls{IP}. Furthermore, \gls{FIB} tables containing forwarding states on \textbf{NDN-LAN} D-switches can leak information for both the consumer and producer. 
Due to the presence of regular \gls{IP} traffic where the communication is not anonymous, \textbf{hICN} guarantees such feature only for \gls{ICN} traffic. Furthermore, the presence of \gls{FIB} tables on hICN-enabled routers might leak information for both consumers and producers. 
Instead, the ICN CSS islands in \textbf{OFELIA} guarantee the anonymity of consumers since their identity is not part of the communication. However, IN and BN might contain information related to end nodes, so the attackers might exploit such information to deanonymize the consumers.

\subsubsection{\textbf{Traffic flow confidentiality}} 
All the analysed hybrid architectures inherit traffic analysis vulnerabilities from both \gls{ICN} and \gls{IP}. In \gls{ICN} islands some vulnerabilities might merge from \gls{ICN} semantics---e.g., human-readable content names. Both types of traffic increase the risk of exposure to this attack. However, padding and other existing countermeasures can be used to prevent these vulnerabilities.

\subsection{Comparison}
\label{ssec:hybrid-comparison}
Table~\ref{tab:hybrid-evaluation} reports the evaluation scores for the analysed hybrid architectures.
Here, \textbf{CLIP} ensures the highest scores. 
\begin{table}[!b] 
\centering
\caption{Evaluation of security and privacy features of hybrid IP-ICN coexistence architectures.}
\label{tab:hybrid-evaluation}
\resizebox{0.55\columnwidth}{!}{
    \begin{tabular}{c | c c c c}
      \toprule
      \makecell{\textbf{SP Feature}} & 
      \makecell{\textbf{\rotatebox{90}{\scriptsize NDN-LAN~\cite{ndnLAN}}}} & 
      \makecell{\textbf{\rotatebox{90}{\scriptsize hICN~\cite{hICN}}}} & 
      \makecell{\textbf{\rotatebox{90}{\scriptsize OFELIA~\cite{ofelia}}}} &
      \makecell{\textbf{\rotatebox{90}{\scriptsize CLIP~\cite{clip}}}} \\
      \toprule 
       \makecell{ Trust} & \tiny \pie{180} & \tiny \pie{360} & \tiny \pie{180} & \tiny \pie{360} \\ 
        \midrule
        \makecell{Data origin \\ authentication} & \tiny \pie{180} & \tiny \pie{0} & \tiny \pie{0} & \tiny \pie{180} \\ 
        \midrule
        \makecell{Peer entity \\ authentication} & \tiny \pie{180} & \tiny \pie{180} & \tiny \pie{180} & \tiny \pie{180} \\ 
        \midrule
        \makecell{Data integrity} & \tiny \pie{180} & \tiny \pie{0} & \tiny \pie{0} & \tiny \pie{180} \\ 
        \midrule
        \makecell{Authorization \\ \& access control}  & \tiny \pie{180} & \tiny \pie{180} &  \tiny \pie{0} & \tiny \pie{180} \\ 
        \midrule
        \makecell{Accountability} & \tiny \pie{180} & \tiny \pie{180} & \tiny \pie{180} & \tiny \pie{180} \\ 
        \midrule
        \makecell{Data \\ confidentiality} & \tiny \pie{0} & \tiny \pie{0} & \tiny \pie{0} & \tiny \pie{0} \\ 
        \midrule
        \makecell{Availability} & \tiny \pie{0} & \tiny \pie{0} & \tiny \pie{0} & \tiny \pie{0} \\ 
        \midrule
        \makecell{Anonymous \\ communication}  & \tiny \pie{0} & \tiny \pie{180} & \tiny \pie{180} & \tiny \pie{0} \\ 
        \midrule
        \makecell{Traffic flow \\ confidentiality} & \tiny \pie{0} & \tiny \pie{0} & \tiny \pie{0} & \tiny \pie{0} \\ 
    \bottomrule
    \end{tabular}}
\\[1\baselineskip]
\tiny \pie{360} \scriptsize Fulfilled~~~~~~ \tiny \pie{180} \scriptsize{Partially fulfilled}~~~~~~\tiny \pie{0} \scriptsize{Not fulfilled}
\end{table}
Indeed, it proposes a data-centric security model based on IPSec and elaborates on three primary security considerations: a) association of the publisher's real identity with the publisher's label, content item label, and the content itself, b) key management, and c) end users' privacy.
After CLIP, \textbf{hICN} scores the highest average. For the design of hICN, the authors made several \gls{SP} considerations on their proposed architecture. Similar to \gls{ICN}, hICN has a data-centric security model, where origin authentication, integrity, and consumer anonymity are enabled by design.
Nevertheless, these architectures fail to ensure data confidentiality, availability, and traffic flow confidentiality. 
Instead, in terms of feature coverage, \textbf{data confidentiality}, \textbf{availability}, \textbf{anonymous communication} and \textbf{traffic flow confidentiality} are amongst the less covered features. Furthermore, \textbf{peer entity authentication} and \textbf{accountability} are the most covered features, mainly due to the presence of IP semantics in hybrid architectures that assure these features.
Lastly, or the hybrid architectures, the presence of both IP and ICN semantics leaves space for existing and new SP challenges. 

\section{Discussion}
\label{sec:discussion}

In this section, we first summarize the findings of this article (Section~\ref{ssec:summary}), and then we discuss the open challenges from this article (\Cref{ssec:openissues}). 
We conclude this section by describing the lessons learned (~\Cref{ssec:lessonslearned}) and the future research directions (Section~\ref{ssec:futuredirections}).

\subsection{Survey summary}
\label{ssec:summary}


This survey aims to provide a comprehensive \gls{SP} evaluation of 20 different coexistence architectures that address the coexistence between \gls{IP} and \gls{ICN} protocols, focusing on design -- i.e., node model and communication scheme -- and implementation particularities.
After that, we analyse each of the described architectures from the \gls{SP} perspective, based on the set of ten selected features. 
We categorised the analysis of the coexistence architectures according to their deployment approach---i.e., overlay, underlay, and hybrid. 
Ten of the analysed architectures adhere to the overlay deployment, while six and four adhere to the underlay and hybrid deployment, respectively.
During the SP analysis of each architecture, we considered the deployment scenarios~\cite{conti2020road} and the additional technologies that some architectures rely on -- e.g., SDN, CDN, DTN, NFV. 
Most architectures aim to connect ICN islands and ICN and IP islands using the current IP infrastructure for the overlay category. This approach is the most efficient and less expensive coexistence scenario since the ICN semantics are integrated into the existing IP infrastructure. Generally, such architectures present a new packet format that encapsulates the ICN semantics in the IP packets or an additional layer that handles mappings between \gls{IP} and \gls{ICN} semantics.
More than 50\% of the analysed overlay architectures -- i.e., NDNFlow, GreenICN, O-ICN, coCONET, OpenFlow-ICN, and SDN-PURSUIT -- exploit SDN functionalities to unlock IP and ICN coexistence. Therefore, during the analysis, we also consider the SP aspects of SDN functionalities that such architectures introduce.
%
Instead, in underlay deployment, \gls{IP} packets are generally translated into \gls{ICN} packets, usually using the gateways. Indeed, this translation process allows the underlay architectures to enable different deployment scenarios while exploiting other technologies. 
Although valid, the presence of gateways or NAP exposes the \gls{ICN} or IP islands beside such nodes to huge risks coming from attackers that, through them, launch prefix hijacking, \gls{DoS} or \gls{DDoS}, and even replay attacks. 
Similar to the underlay category, the architectures adhering to the hybrid deployment approach offer a wide range of deployment scenarios. Indeed, dual-stack routers that process IP and ICN traffic enable communication between different islands over IP and ICN oceans.
\Cref{tab:summary} summarizes the findings of this article, mainly represented in \Cref{tab:overlay-evaluation,tab:underlay-evaluation,tab:hybrid-evaluation}.
\begin{table*}[!ht] 
\centering
\caption{Summary of the addressed SP features during architecture design. The assigned number corresponds to the number of architectures in each deployment approach that address the corresponding feature in the row -- i.e., extracted by counting fully and partially fulfilled features from \Cref{tab:overlay-evaluation,tab:underlay-evaluation,tab:hybrid-evaluation} -- out of the total number of the analysed architectures corresponding to such deployment---i.e., 10 overlay, 6 underlay and 4 hybrid architectures.}
\label{tab:summary}
    \begin{tabular}{ c  c  c  c  c  c  c }
      \toprule
       \makecell{\multirow{3}{*}{\textbf{SP features}}} &
       \multicolumn{2}{c}{\makecell{\textbf{Overlay}}} & \multicolumn{2}{c}{\makecell{\textbf{Underlay}}} & \multicolumn{2}{c}{\makecell{\textbf{Hybrid}}} \\
       \cline{2-7} \\
       & \makecell{\textbf{Fully} \\ \textbf{fulfilled}} & \makecell{\textbf{Partially} \\ \textbf{fulfilled}} & \makecell{\textbf{Fully} \\ \textbf{fulfilled}} & \makecell{\textbf{Partially} \\ \textbf{fulfilled}} & \makecell{\textbf{Fully} \\ \textbf{fulfilled}} & \makecell{\textbf{Partially} \\ \textbf{fulfilled}} \\
      \toprule 
       \makecell{Trust} & $\frac{2}{10}$ & $\frac{8}{10}$ & $\frac{1}{6}$ & $\frac{5}{6}$ & $\frac{2}{4}$ & $\frac{2}{4}$  \\
        \midrule
        \makecell{Data origin \\authentication} & $\frac{1}{10}$ & $\frac{9}{10}$ & $\frac{1}{6}$ & $\frac{1}{6}$ & 0 & $\frac{2}{4}$ \\
        \midrule
        \makecell{Peer entity \\authentication} & 0 & $\frac{3}{10}$ & 0 & $\frac{2}{6}$ & 0 & $\frac{4}{4}$ \\
        \midrule
        \makecell{Data \\ integrity} & $\frac{1}{10}$ & $\frac{9}{10}$ & $\frac{1}{6}$ & $\frac{1}{6}$ & 0 & $\frac{2}{4}$ \\
        \midrule
        \makecell{Authorization \& \\ access control} & 0 & $\frac{6}{10}$ & 0 & $\frac{4}{6}$ & 0 & $\frac{3}{4}$ \\
        \midrule
        \makecell{Accountability} & 0 & $\frac{10}{10}$ & $\frac{1}{6}$ & $\frac{5}{6}$ & 0 & $\frac{4}{4}$ \\
        \midrule
        \makecell{Data \\ confidentiality} & 0 & 0 & $\frac{1}{6}$ & 0 & 0 & 0 \\
        \midrule
        \makecell{Availability} & 0 & $\frac{1}{10}$ & 0 & 0 & 0 & 0 \\
        \midrule
        \makecell{Anonymous \\ communication} & $\frac{1}{10}$ & $\frac{5}{10}$ & 0 & $\frac{2}{6}$ & 0 & $\frac{2}{4}$ \\
        \midrule
        \makecell{Traffic flow \\ confidentiality} & 0 & 0 & 0 & $\frac{1}{6}$ & 0 & 0 \\
        
    \bottomrule
    \end{tabular}
\end{table*}
Overall, almost all the coexistence architectures at least partially fulfill data origin and peer entity authentication, data integrity, authorization and access control, and accountability. On the contrary, they face challenges in supporting the other features. 
Contrarily, only a few architectures fully fulfill the SP requirements. Generally, such architectures explicitly and clearly describe SP mechanisms to achieve the analysed features during the design phase. 
Taking into account the analysis scores, the deployment scenarios, and the other technologies, we summarize some considerations for the deployment categories as follows:
\begin{description}
    \item[\textbf{Overlay.}] Almost none of the architectures, except coCONET~\cite{coCONET}, explicitly address SP considerations during the design phase. Indeed, the coexistence scenarios that include heterogeneous communication models must also properly design their security model.
    Given the deployment nature of overlay category -- i.e., encapsulation of ICN in IP~\cite{fahrianto2020comparison} -- relatively all features are at least partially fulfilled. 
    Indeed, data origin authentication and integrity for the ICN islands are partially fulfilled since only the data packet signature can be verified. At the same time, this rule does not apply to interest packets. 
    Additionally, the communication among islands is based on IP packets, allowing for establishing accountability and authorization and access control procedures whenever the IP addresses are present. The SDN-based architectures can use the control plane to establish access control procedures for the latter.
    Lastly, the analysed overlay architectures almost entirely lack peer entity authentication, data confidentiality, availability, anonymous communication, and traffic flow confidentiality.
    
    \item[\textbf{Underlay.}] From the analysed underlay architectures, COIN~\cite{coin} is the only architecture that designs a proper trust model, fully describing the encryption and signature mechanisms. The remaining architectures at least partially fulfill features such as trust, data origin authentication and integrity, authorization and access control, and accountability. 
    On the contrary, achieving peer entity authentication, data confidentiality, availability, anonymous communication, and traffic flow confidentiality is challenging for underlay deployment. This is mainly connected to the presence of different scenarios. 

    \item[\textbf{Hybrid.}] hICN and NDN-LAN introduce the dual-stack switches among the four analyzed hybrid architectures. Although an expensive approach~\cite{fahrianto2020comparison}, it enables ICN benefits -- e.g., in-network caching, name-based forwarding -- in the existing IP infrastructure. 
    Like the other two categories, hybrid architectures fail to ensure data confidentiality, availability, anonymous communication, and traffic flow confidentiality.
    
\end{description}

\subsection{Open issues}
\label{ssec:openissues}



Following the findings presented in Table~\ref{tab:summary} and the considerations of~\Cref{ssec:summary}, data confidentiality, availability, anonymous communication and traffic flow confidentiality are four features that most architectures can not fully achieve. Believing that these features remain an open challenge in deploying secure \gls{IP}-\gls{ICN} coexistence architectures, we subsequently provide some considerations for each.   
\begin{description}
    \item[\textbf{Data confidentiality.}] The majority of the analyzed architectures do not elaborate on the mechanisms that can be used to ensure data confidentiality. Achieving such features in architectures that combine different domains is more challenging due to different encryption schemes and key management infrastructures. COIN is the only architecture that transparently describes the gateway importance during the design. COIN gateway decrypts the data from one domain and re-encrypts them for the other domain. Furthermore, the NDN testbed we have analyzed partially ensures this feature.
    Generally, to establish the connection between \gls{IP} and \gls{ICN} domains, IP addresses are translated to name prefixes and vice versa. Even if IP headers do not reveal information that might compromise data confidentiality, name prefixes might disclose information about the content. The ICN islands can use existing data confidentiality solutions such as attribute-based schemes~\cite{ion2013toward}.
    
    \item[\textbf{Availability.}] According to the analysis carried out in this article, availability is one of the most difficult \gls{SP} features to be achieved in both clean state \gls{ICN} and \gls{IP}-\gls{ICN} coexistence architectures. \gls{DoS} and \gls{DDoS} are well-known attacks in \gls{IP} networks that target network availability, and it is demonstrated that such attacks are present also in \gls{ICN} networks~\cite{zhang2013caching}. Due to added features -- i.e., the presence of \gls{PIT} tables and caches in \gls{ICN} routers -- different attacks are targeting the routers. Such attacks include interest flooding~\cite{afanasyev2013interest, compagno2013poseidon}, content poisoning~\cite{ghali2014network} and cache pollution~\cite{conti2013lightweight,man2021cache}. In this context, the coexistence architectures inherit such attacks in both \gls{IP} and \gls{ICN} islands. Additionally, the underlay approaches usually introduce \gls{NAP} nodes which malicious users can target to launch prefix hijacking and replay attacks~\cite{conti2020road}. Furthermore, underlay architectures use gateways that enable the translation of information from one domain to another. Such nodes can also be a target for resource exhaustion attacks that degrade their availability. Some architectures -- e.g., PURSUIT, POINT, RIFE -- maintain states in \gls{RF} and \gls{TF} functions. Attackers can exploit these functions to cause the introduction of new states and easily disrupt the availability. Most coexistence architectures -- e.g., NDNFlow, O-ICN, GreenICN, SDN-PURSUIT, POINT, RIFE, and OpenFlow-ICN -- exploit the \gls{SDN} concept. Here, the controller is considered a single point of failure since its failure affects the entire network and exposes it to risks related to availability. An exception is made for coCONET, which requires each ICN node to verify the signature before forwarding the content. Nevertheless, as the authors confirmed, such a solution is not lightweight.
    
    \item[\textbf{Anonymous communication.}] The endorsement of this feature depends on the deployment scenario. In this context, in \gls{IP} islands, the communication is still established using the \gls{IP} addresses. Therefore, the communicating hosts can not remain anonymous during communication. Conversely, in \gls{ICN} islands, the communication is established by using the content names, so the anonymity of participating hosts is ensured. In this context, the anonymous communication feature is partially fulfilled whenever the ICN islands are present.
    However, for evaluating this feature, we also encountered the presence of elements that store information regarding nodes participating in the network, which attackers might maliciously exploit to deanonymise them.
    To mitigate the lack of communication anonymity in \gls{IP} islands, systems such as The Onion Router (TOR)~\cite{tor} can be used. Nevertheless, several works have demonstrated that Tor is still not anonymous enough~\cite{basyoni2020traffic}. Furthermore, some works have been proposed to ensure producer's anonymity in ICN networks~\cite{al2022harpocrates,kita2022private,dibenedetto2011andana,tsudik2016ac3n}.
        
    \item[\textbf{Traffic flow confidentiality.}] The researchers have proven that encryption is not enough to avoid information leakage from traffic flow in \gls{IP} networks~\cite{raymond2001traffic}. Some countermeasures against traffic analysis have been proposed, such as TOR~\cite{tor} or traffic morphing~\cite{wright2009traffic}. Nevertheless, several works~\cite{dyer2012peek, cai2012touching, panchenko2016website} showed that these countermeasures still fail when coarse-grained side-channel attacks are applied, or the attack training set is accurately selected. 
    Similarly, \gls{ICN} traffic can be susceptible to traffic analysis~\cite{ambrosin2018security}. Therefore, the coexistence architectures must assess the issues the research community raises.
\end{description}

\subsection{Lessons learned}
\label{ssec:lessonslearned}
%
%
Throughout the analysis carried out in this paper, we showed that the IP-ICN coexistence has principally been studied only from the deployment and performance point of view~\cite{conti2020road,fahrianto2020comparison}. Even though the current IP history taught us the importance of addressing security and privacy requirements beforehand, considering them when designing new architectures is still not a good practice. Indeed, in a coexistence scenario where heterogeneous protocols are considered, and each has a different security model, the SP requirements must be regarded appropriately during the design. As we showed in this article, only a few proposals address this issue.
Additionally, ensuring most SP features is not trivial in an IP-ICN coexistence scenario. Here, host- and content-based security models must be adequately integrated and mapped to ensure the robustness and trustworthiness of the coexistence architectures. Even for these requirements, only a few architectures handle them properly.  
The article illustrates that multiple architectures exploit other emerging technologies to enable IP-ICN cohabiting. Although efficient, these technologies transfer known SP issues in the coexistence scenario. Therefore, a trade-off between the performance and security of the designed coexistence architectures must be considered during design.

\subsection{Future direction}
\label{ssec:futuredirections}
Considering the above considerations on the lessons learned in this article, we present some directions that need further investigation in this research field.
\begin{description}
    \item[\textbf{Selection of a secure coexistence approach.}] 
    Securing the IP-ICN coexistence is indeed a difficult task. This article showed that the current state-of-the-art encompasses only a few secure-by-design architectures. 
    Generally, the overlay architectures that introduce a new packet format that allows the mapping of ICN faces into \gls{IP} addresses ensure to trade on security features from the IPSec suite.
    Although easily implementable in the current infrastructure, tunneling IP channels to enable ICN communication does not permit these architectures to benefit from ICN advantages---e.g., caching on routers and name-based content retrieval.  
    In the future, providing more research on enabling ICN features on overlay-like infrastructures wisely can help overcome the caching issue and place the content near the users. In particular, we believe that the ICN-based IoT networks benefit from efficient cache placement policies as IoT devices are resource-constrained~\cite{alduayji2023pf}.
    On the contrary, the underlay and hybrid architectures empower the ICN features. 
    Here, the former architectures introduce new nodes -- e.g., NAP, gateways -- important for translating from \gls{IP} to \gls{ICN} semantics and vice versa. However, from the \gls{SP} point of view, these newly introduced entities can be a target of different attacks, mostly related to \gls{DoS} attacks. 
    In this approach, the translation nodes must also be able to translate from one type of security model -- \gls{IP} or \gls{ICN} -- to the other, introducing further complexity in such nodes. 
    In this perspective, future research must target the design of efficient and lightweight mapping mechanisms from the IP to the ICN security model and vice-versa. Here, the gateways can be leveraged as a potential point to deploy such mechanisms.
    Lastly, hybrid architectures generally accommodate both types of traffic through the presence of dual-stack nodes that can route both IP and ICN packets.
    The presence of both types of traffic requires these architectures to consider the involvement of both types of security models -- i.e., host-based and content-based -- and enable them during design. 
    In the future, new intrusion detection and mitigation mechanisms that encompass both \gls{IP} and \gls{ICN} known attacks can contribute to enhancing security while being deployed on dual-stack network devices.
    %
    \item[\textbf{Selection of secure additional technologies.}] \gls{SDN}, \gls{NFV}, \gls{DTN}, and \gls{CDN} are some of the technologies used by several coexistence architectures to improve the coexistence between \gls{IP} and \gls{ICN}. 
    In particular, most of the surveyed architectures exploit \gls{SDN} technology, separating the data from the control plane.
    Here, SDN simplifies the deployment of \gls{ICN} communication through the controller that manages and installs \gls{ICN} rules in the switches. 
    However, from the \gls{SP} perspective, the controller introduces a single point of failure in the network that can be the target of different attacks. Furthermore, the architectures that use the \gls{CDN} technology where the presence of caches is distributed all over the network have to face the availability issues that might arise. 
    In this context, further investigation is needed to assess and evaluate the integration of other technologies, particularly on the \gls{SP} aspects.
    We believe there is no silver bullet solution regarding selecting an additional technology to be used, and a trade-off must be made between the benefits obtained in applying technology and the \gls{SP} issues it introduces.
    \item[\textbf{Securing the proposed architectures.}]
    Besides selecting the secure coexistence approach and additional technologies, future dedicated research can yield more secure and private the proposed coexistence architectures.
    To this end, in the future, open source IP-ICN coexistence implementations -- i.e., NDN, CONET, NDNFlow, POINT --, simulators -- i.e., O-CIN -- and hardware prototypes -- i.e., hICN switch -- can be used to experiment and make the proposed architectures more robust.
    %
\end{description}

\section{Conclusions}
\label{sec:conclusions}

In this article, we surveyed \gls{IP}-\gls{ICN} coexistence architectures proposed by both Academia and Industry while analyzing their security and privacy aspects. 
The focus of these proposals has mainly been to exploit some benefits of \gls{ICN} -- i.e., forwarding, storage, and security -- and integrate them into existing \gls{IP} infrastructure. 
None of the existing proposals evaluate the security and privacy costs deriving from the combination of heterogeneous protocols. 
Thus, this article presents the first effort to study these proposals from the security and privacy point of view while pointing out that the path towards future Internet must include a secure coexistence phase between the old and future protocols. Our analysis is based on a large set of security and privacy features which allowed us to identify that most proposed architectures present gaps in providing security and privacy procedures. A secure coexistence phase can be ensured by balancing network performance goals -- i.e., low latency, improved throughput -- and security procedures that provide secure communication. 
In particular, the surveyed coexistence architectures can be improved, especially in four security and privacy features: data confidentiality, availability, anonymous communication, and traffic flow confidentiality. Furthermore, the use of other technologies that enable the IP-ICN coexistence must be adequately considered not only from the performance point of view but also for the security and privacy issues that might introduce in the coexistence scenario.
Concluding, we firmly believe that both the future directions and open issues we pointed out shed some light on eventual motivations for researchers to investigate more towards secure \gls{IP}-\gls{ICN} coexistence. 

\balance
\bibliographystyle{IEEEtran}
\bibliography{ref}



 





\end{document}